\documentclass[useAMS,usenatbib]{mn2e}

\pdfoutput=1

\usepackage{graphicx}
\usepackage{xtab,afterpage}
\usepackage{amsmath}
\usepackage{epstopdf}
\usepackage{url}
\usepackage[flushleft]{threeparttable}
\usepackage{booktabs,fixltx2e}
\usepackage{amssymb}

\newcommand{\rmxaa}{Revista Mexicana de Astronomia y Astrofisica}

\newcommand{\pasp}{{\it PASP~\/}}

\newcommand{\apj}{ApJ}
\newcommand{\apjl}{ApJ}
\newcommand{\apjs}{ApJS}
\newcommand{\aap}{A \& A}
\newcommand{\araa}{ARA\&A}
\newcommand{\aj}{AJ}
\newcommand{\mnras}{MNRAS}

\newcommand{\nat}{Nature}

\def\apss{\rmfamily{Ap\&SS~}}     

\makeatletter
\newcommand*{\rom}[1]{\expandafter\@slowromancap\romannumeral #1@}
\makeatother

\title[Icy exocomets as the origin of gas in debris discs]{Predictions for the secondary CO, C and O gas content of debris discs from the destruction of volatile-rich planetesimals}
\author[Q. Kral]{Quentin Kral\thanks{E-mail: qkral@ast.cam.ac.uk}, Luca Matr\`a, Mark C. Wyatt, Grant M. Kennedy \\
Institute of Astronomy, University of Cambridge, Madingley Road, Cambridge CB3 0HA, UK}

\begin{document}

\date{Accepted 1928 December 15. Received 1928 December 14; in original form 1928 October 11}

\pagerange{\pageref{firstpage}--\pageref{lastpage}} \pubyear{2002}

\maketitle

\label{firstpage}

\begin{abstract}
This paper uses observations of dusty debris discs, including a growing number of gas detections in these systems, to test our understanding of the origin and evolution of this gaseous component. It is assumed that all debris discs with icy planetesimals create second generation 
CO, C and O gas at some level, and the aim of this paper is to predict that level and assess its observability. We present a new semi-analytical equivalent of the numerical model of \citet{2016MNRAS.461..845K} allowing application to large numbers of systems. That model 
assumes CO is produced from volatile-rich solid bodies at a rate that can be predicted from the debris disc’s fractional luminosity. CO photodissociates rapidly into C and O that then evolve by viscous spreading. This model provides a good qualitative explanation of all current 
observations, with a few exceptional systems that likely have primordial gas. The radial location of the debris and stellar luminosity explain some non-detections, e.g. close-in debris (like HD 172555) is too warm to retain CO, while high stellar luminosities (like $\eta$ Tel) result 
in short CO lifetimes. We list the most promising targets for gas detections, predicting $>15$ CO detections and $>30$ CI detections with ALMA, and tens of CII and OI detections with future far-IR missions. We find that CO, CI, CII and OI gas should be modelled in non-LTE for 
most stars, and that CO, CI and OI lines will be optically thick for the most gas-rich systems. Finally, we find that radiation pressure, which can blow out CI around early-type stars, can be suppressed by self-shielding.

\end{abstract}


\section{Introduction}

Gas is observed around a growing number of planetary systems where planets are likely to be formed and the protoplanetary discs in which they formed are already gone. 
All these gas detections are in systems where secondary dust is created from collisions by bigger bodies
orbiting in a debris belt similar to the Kuiper or asteroid belt in our solar system. Similarly, the observed gas around these mature systems 
may also be of secondary origin and being released from debris belt planetesimals/dust owing to grain-grain collisions \citep{2007ApJ...660.1541C}, planetesimal breakup \citep{2012ApJ...758...77Z}, sublimation \citep[e.g.][]{1990A&A...236..202B}, photodesorption \citep{2007A&A...475..755G}
or giant impacts \citep{2009ApJ...701.2019L,2014MNRAS.440.3757J}. For some systems such as HD 21997, the observed gas may be of primordial origin \citep{2013ApJ...776...77K}.

Molecular CO gas is observed in the sub-mm with both single-dish telescopes (JCMT, APEX) and interferometers such as ALMA, the SMA or NOEMA. For the brightest targets, ALMA's high-resolution and unprecedented sensitivity
allow us to obtain CO maps for different lines and isotopes showing the location of CO belts and giving an estimate of their mass 
\citep[see the CO gas disc around $\beta$ Pic,][]{2014Sci...343.1490D,2016MNRAS}. Atomic species are also detected around a few debris disc stars. In particular, Herschel was able to detect
the OI and CII fine structure lines in two and four systems, respectively \citep[e.g.][]{2012A&A...546L...8R,2013ApJ...771...69R,2014A&A...565A..68R,2014A&A...563A..66C,2016A&A...591A..27B}. Also, metals have been detected, using UV/optical absorption lines,
around $\beta$ Pictoris \citep[Na, Mg, Al, Si, and others,][]{2006Natur.441..724R}, 49 Ceti \citep[CaII,][]{2012PASP..124.1042M}, and HD 32297 \citep[NaI,][]{2007ApJ...656L..97R}. Some of these metals are on Keplerian orbits but should be blown out by the ambient radiation pressure \citep{2001ApJ...563L..77O}.
It is proposed that the overabundant ionised carbon observed around $\beta$ Pic, which is not pushed by radiation pressure could brake other ionised species due to Coulomb collisions with them \citep{2006ApJ...643..509F}.
A stable disc of hydrogen has not yet been observed in these systems \citep{1995A&A...301..231F,2001Natur.412..706L} but some high velocity HI component (presumably falling onto the star) was detected recently with the HST/COS around $\beta$ Pic \citep{2016arXiv161200848W}. 
All these observations need to be understood within the framework of a self-consistent model. Models of the emission of the gas around main sequence stars have been developed, but gas radial profiles were not derived self-consistently and often assumed
to be gaussian \citep[e.g.,][]{2010ApJ...720..923Z} or not to be depleted in hydrogen compared to solar \citep[as expected in debris discs, e.g.,][]{2004ApJ...613..424G} or both \citep[e.g.,][]{2000A&A...353..276K}.

One self-consistent model has been proposed in \citet{2016MNRAS.461..845K} (KWC16) that can explain gas observations around $\beta$ Pictoris. 
It proposes that CO gas is released from solid volatile-rich bodies orbiting in a debris belt as first proposed by \citet{2011ApJ...740L...7M,2012ApJ...758...77Z}, and verified by \citet{2014Sci...343.1490D,2016MNRAS}. CO is then photodissociated quickly and produces atomic carbon and oxygen gas
that evolves by viscous spreading, parameterised with an $\alpha$ viscosity, resulting in an accretion disc inside the parent belt and a decretion disc outside. A steady state is rapidly reached (on a viscous timescale), meaning that it is unlikely that we observe a system in a transient phase.
The $\alpha$ viscosity could be provided by the magneto-rotational instability (MRI) as presented in \citet{2016MNRAS.461.1614K}. Gas temperature, ionization state and population levels are computed using
the photodissociation region (PDR) model Cloudy at each time step \citep{2013RMxAA..49..137F}. 

This model is generic and could apply to all debris discs as long as they are made up of volatile-rich bodies and their CO content is released as gas as they are ground down within a steady-state collisional cascade as proposed in \citet{2015MNRAS.447.3936M}. 
To apply the KWC16 model to a given system, we assume that the CO released is a proportion $\gamma$ of the mass lost through the collisional cascade. That mass loss rate can be determined from the fractional luminosity of the debris disc $L_{\rm IR}/L_\star$ and its temperature from which the 
planetesimal belt location $R_0$ can be determined. These combine to give the CO input rate of $\dot{M}_{\rm CO}$, which is one of the parameters in the KWC16 model, along with $R_0$, the $\alpha$ viscosity, the amount of radiation coming from the 
central star (L$_\star$) and the interstellar radiation field ($L_{\rm IRF}$) as well as the distance to Earth $d$. That model can then provide as an output the radial structure of the atomic gas disc in terms of density, temperature, ionization fraction for different elements, line fluxes,
and make predictions/produce synthetic images for observations with ALMA or any other instruments.

In this paper, we will provide a semi-analytical model (simpler than the complex numerical model described above) to model secondary gas in debris discs and apply it
to a large sample of debris disc systems in order to predict the abundance and detectability of CO, CI, CII and OI for each system. Previous observations, both detections and non-detections, will provide tests of these predictions, 
and allow us to assess whether the model can be used as a reliable predictor for unobserved systems.

We assume $\gamma$, $\alpha$, and $L_{\rm IRF}$ are same for all stars \citep[we take the local interstellar radiation field derived by][]{2011piim.book.....D}, in which case the detectability of gas in the model depends only on $R_0$, $\dot{M}_{\rm CO}$, L$_\star$ and $d$. 
We will explain which parts of this parameter space should be preferentially observed when trying to detect CO, CI, CII or OI with different instruments. This will give a general understanding of gas observations in debris discs and is particularly
well suited for planning mm-wave APEX/ALMA line observations and considering the science that could be done with future missions such as SPICA \citep{2009ExA....23..193S} or NASA's far-IR surveyor concept (FIRS, now called the Origins Survey Telescope) 
that may be built within the next 15 years.

In section \ref{gasobs}, a summary of gas detections around nearby main sequence stars is presented. In section \ref{CO}, we present CO abundance predictions for a large sample of debris discs and explain which systems are more likely to have CO detected. In section \ref{carbon}, we present
a similar analysis for CI and CII and go on with OI predictions in section \ref{ox}. We discuss our findings in section \ref{disc} before concluding in section \ref{ccl}.

\section{Gas observations in debris discs}\label{gasobs}

\begin{figure}
   \centering
  \includegraphics[width=9.5cm]{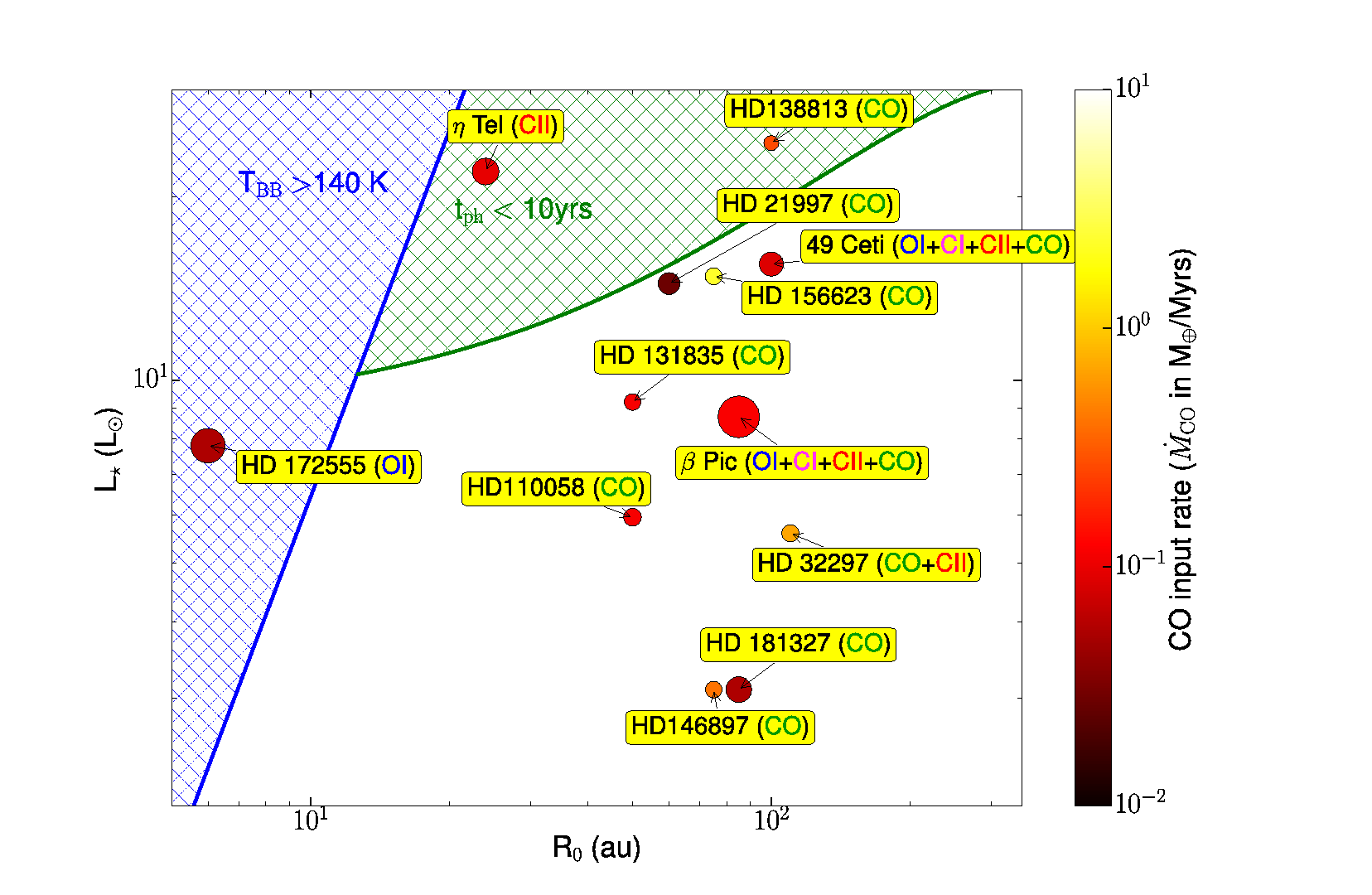}
   \caption{\label{fig1} L$_\star$ Vs $R_0$ for all discs with gas detected (see Table \ref{tab1}). The point sizes are inversely proportional to distance to Earth and the colour bar indicates CO input rates deduced from our model. The blue line represents a black body temperature of 140K and the 
green line a photodissociation timescale of 10 years.}
\end{figure}

The number of debris disc systems with gas detected is growing and we now have 12 systems that can help us to understand the dynamics of this gas and its origin. These systems are presented in Table \ref{tab1}. Ten of them have CO detections, whilst 2 have CI detected, 
4 have CII detected and 3 systems have OI detected.
All of these systems are shown in Fig.~\ref{fig1} in a $L_\star$ versus $R_0$ diagram (see Table \ref{tab1} to find the values used and their references). 

In this plot, we show the 4 fundamental parameters that matter in this study: $R_0$ and L$_\star$ (the x and y-axes), $\dot{M}_{\rm CO}$\footnote{Computed with Eq.~\ref{COloss}} (the point colour)
, and $d$ (the point size). On the plot, one can see where the 12 systems lie, and we annotate their names, as well as the elements that have been detected so far \citep[we omit metals as they are not expected to make up the bulk of the gas,][]{2006ApJ...643..509F}. 
We overlay a blue line showing a black body temperature of 140K. For CO adsorbed on amorphous H$_2$O, this is the temperature above which CO cannot be trapped in ices \citep[under laboratory conditions,][]{2003ApJ...583.1058C}. Any systems to the left 
of this line should not be able to retain any CO on grains (if no refractories are present to hold CO). A photodissociation timescale of 10 years is shown by the green line. If only the interstellar radiation field (IRF) were to be present (no central star or shielding), the photodissociation timescale would be equal to $\sim $120 years \citep{2009A&A...503..323V}. Systems that are above
the green line are sufficiently luminous that the central star's radiation will act to significantly decrease this timescale. It is thus unlikely to detect CO far above this green line for reasonable production rates.

The CII and OI fine structure lines have been detected with Herschel thanks to the GASPS program, which used PACS to survey a small sample of debris discs \citep{2013PASP..125..477D,2014A&A...565A..68R}. Also, HIFI high spectral resolution data 
have been used for probing the CII location (and potential asymmetries) around $\beta$ Pic \citep{2014A&A...563A..66C}. Of the systems with CO detected, 9 out of 10 have also been observed and detected with ALMA (see Table~\ref{tab4} to see the calculated CO masses from line fluxes). We included the four new CO ALMA detections by \citet{2016ApJ...828...25L}, 
HD 138813, HD 146897, HD 110058 and HD 156623 (references are listed in Table~\ref{tab1}). We do not include the recent tentative detection of CO in $\eta$ Corvi by \citet{2016arXiv161101168M}, as the CO detection is not co-located with its debris belt and the gas release mechanism may be different than proposed in this paper. 

\begin{table*}
    \caption{List of all known debris discs with gas.}
\begin{center}

\begin{threeparttable}
\begin{tabular}{|l|c|c|c|c|c|c|c|c|}
  \toprule
  Star's  & Atoms \& & L$_\star$ & $d$ & $M_{\rm CO}$ & $R_0$ & M$_{\rm dust}$ & $L_{\rm IR}$/$L_\star$ & Star's\\
  name & molecules observed & (L$_\odot$) & (pc) & (M$_\oplus$)$^a$ & (au) & (M$_\oplus$) & & age (Myr)\\
  \midrule
  $\beta$ Pic$^{1}$ & CO,CI,CII,OI,...&8.7&19.4&$2.8\times10^{-5}$&85& $7.8\times10^{-2}$ & $1.7 \times 10^{-3}$ & 23\\
  49 Ceti$^{2}$ & CO,CI,CII,OI$^b$&15.5&59.4&$1.4\times10^{-4}$&100& 0.27 & $1.1 \times 10^{-3}$ & 40\\
  $\eta$ Tel$^{3}$ & CII&22&48.2&-&24& $1.3\times10^{-2}$ & $7.6 \times 10^{-4}$ & 23\\
  HD 21997$^{4}$ & CO&14.4&71.9&$6\times10^{-2}$&60& 0.16 & $5.9 \times 10^{-4}$ & 45\\
  HD 32297$^{5}$ & CO,CII&5.6&112&$1.3\times10^{-3}$&110& 0.37 & $5.4 \times 10^{-3}$ & 30\\
  HD 110058$^{6}$ & CO&5.9&107&$2.1\times10^{-5}$&50& $3\times10^{-3}$ & $1.9 \times 10^{-3}$ & 10\\
  HD 131835$^{7}$ & CO&9.2&122&$6\times10^{-2}$&50& 0.47 & $1.5 \times 10^{-3}$ & 16\\
  HD 138813$^{6}$ & CO&24.5&150.8&$7.4\times10^{-4}$&100& $7.6\times10^{-3}$& $1.5 \times 10^{-3}$ & 10\\
  HD 146897$^{6}$ & CO&3.1&122.7&$2.1\times10^{-4}$&100& $2\times10^{-2}$ & $5.4 \times 10^{-3}$ & 10\\
  HD 156623$^{6}$ & CO&14.8&118&$2.0\times10^{-3}$&75& $2.4\times10^{-4}$ & $5.5 \times 10^{-3}$ & 10\\
  HD 172555$^{8}$ & OI&7.8&29&-&6& $4.8\times10^{-4}$ & $7.8 \times 10^{-4}$ & 23\\
  HD 181327$^{9}$ & CO&3.1&51.8&$1.8\times10^{-6}$&85& 0.44 & $2 \times 10^{-3}$ & 23\\
  \bottomrule
\label{tab1}
\end{tabular}
\begin{tablenotes}
		\footnotesize
		\item $^a$ We computed these masses in NLTE from the most recent $^{12}$CO integrated line fluxes cited in papers below (assuming that it is optically thin), except for HD 21997 and HD 131835 where it is based on C$^{18}$O observations (see also Table~\ref{tab4}).
		\item $^b$ CI and OI are detected via absorption lines in the UV for 49 Ceti and their abundances are not well quantified \citep{2014ApJ...796L..11R}.
   		\item $^1$ \citet{2016MNRAS.461..845K}, $^2$ \citet{2008ApJ...681..626H},$^3$ \citet{2014A&A...565A..68R},$^4$ \citet{2013ApJ...776...77K},$^5$ \citet{2016MNRAS.461.3910G},$^6$ \citet{2016ApJ...828...25L},$^7$ \citet{2015ApJ...814...42M,2016conf},$^8$ \citet{2012A&A...546L...8R},$^9$ \citet{2016MNRAS.460.2933M}
  	\end{tablenotes}
\end{threeparttable}

\end{center}

\end{table*}



In Fig.~\ref{fig1} and Table~\ref{tab1}, one can notice several trends. Most gas detections are around A stars (only HD 181327 and HD 146897 are F stars). Also, all systems have high fractional luminosities; greater than $5 \times 10^{-4}$. Moreover, all the detections are for young systems that are less 
than 45 Myr old\footnote{We note that the age of HD 32297 is not well constrained but is likely $\sim$ 30Myr or younger \citep{2005ApJ...635L.169K}.}. In terms
of the gaseous species detected, CO is almost always detected (10/12). For one system, CII is detected without CO ($\eta$ Tel), and for another, OI is the only element detected (HD 172555). 
These two systems are located in the green and blue hatched areas, respectively, which could potentially 
explain a lack of CO detections so far (see subsections \ref{firstc} and \ref{detecta} for a more thorough explanation).
We also note that there is an OI detection around HD 98800 \citep[member of TW Hydrae association,][]{2013A&A...555A..67R} but we do not include
this system as it might still be in an early pre-debris disc stage.
All of the CO detections are for systems with debris belts located beyond 50au. 

Could these main trends be explained within the framework presented in KWC16? The remainder of this paper
tackles this question. We first present the part of the semi-analytical model that computes CO mass predictions from the parameters of the dust belt and all the results we get for CO in section \ref{CO}. We then present the rest of the semi-analytical
model to be able to get CI, CII predictions in section \ref{carbon} and later OI predictions in section \ref{ox}. We show that it is indeed possible to explain all the main trends presented above and henceforth give some predictions for debris disc systems without gas detected so far.

\section{Understanding CO}\label{CO}

In this section, we explain why CO has been detected only around 10 main sequence stars so far. To do so, we use our model to make predictions for the CO mass around many debris discs under the assumption that the dust is created in the destruction of volatile-rich planetesimals, 
a process which also releases CO gas \citep{2012ApJ...758...77Z,2015MNRAS.447.3936M}.
We then compare these predictions to APEX and ALMA mass detection limits to assess the detectability of each system. We then make predictions of CO detectability around a large number of debris disc host stars and provide the most promising targets 
to observe in the near future. We also identify what determines the abundance of CO in any system. We show how observations give us a way to access the CO content of planetesimals, from which the observed CO is released \citep{2015MNRAS.447.3936M,2016MNRAS.460.2933M,2016MNRAS}.

\subsection{First check: Solid body temperature and photodissociation timescale}\label{firstc}

Fig.~\ref{fig1} gives some first insights on the systems in which CO is most likely to be detected. If the system is located in the blue hatched area, all CO has likely been lost already as it is released from icy grains above $T\sim$140K. This conclusion assumes that there are no refractories and no CO
hidden in the core of big rocky bodies (the blue line assumes that grains radiate like black bodies). The only possibility to
have CO in this region in a secondary scenario would be if there is enough CO \citep[or CI,][]{2016MNRAS} to shield the radiation coming from both the star and IRF, but this requires a substantial amount \citep{2016MNRAS}. This explains naturally
why we do not expect CO to be detected around HD 172555 but we note that if the disc is twice as large as assumed here and/or that CO is hidden inside rocky bodies and released when they collide, it would be possible for CO to be present. 

Also, it is less likely to find CO in the green hatched area as the photodissociation timescale is smaller than 10 years (calculated with Eq.~\ref{phodi}, assuming no shielding) in this region and can reach very low values. This is a natural explanation for the lack of CO detection around $\eta$ Tel so far.
Not accounted for in this explanation are the CO mass input rate $\dot{M}_{\rm CO}$ and distance to Earth $d$, which we consider further below.

\subsection{CO mass predictions}

To predict the CO mass within debris disc systems, we make the assumption that gas is produced from debris created through the collisional cascade. The mass loss rate can be worked out from \citet{2008ARA&A..46..339W} assuming a $q=-3.5$ standard size distribution \citep[e.g.][]{2013A&A...558A.121K}. While producing debris through
the collisional cascade, we assume that solid bodies are composed\footnote{Note that the CO$_2$ ice may also contribute to the observed CO gas mass \citep[e.g.][]{2016MNRAS.460.2933M}, in which case this assumed CO+CO$_2$ fraction would be higher by at most a factor of a few,
increasing the CO gas mass produced (this would only change $\gamma$ by a few).} of a certain amount of CO $\gamma$ \citep[typically equal to $\sim$ 10 percent in Solar System comets,][]{2011ARA&A..49..471M} that is released through the mass loss process. Solid bodies ground down into dust
in the collisional cascade are removed by radiation pressure at a rate $\dot{M}_{\rm loss}$. Unless volatiles remain in dust then the CO production rate should only depend on the rate of dust production. In terms of the parameter space we study in this paper, the mass loss rate is equal to \citep{2008ARA&A..46..339W}

\begin{equation}
\label{massloss}
\dot{M}_{\rm loss}=\delta \left( \frac{L_{\rm IR}}{L_\star} \right)^2 \left( \frac{L_\star}{L_\odot} \right)^{13/12} \left( \frac{R_0}{1 {\rm au}} \right)^{-1/3} {\rm M}_\oplus/{\rm Myr},
\end{equation}

\noindent where $R_0$ is the distance from the host star to the planetesimal belt (in au), $L_\star$ is the star's luminosity (in L$_\odot$) and $\delta$ equals $e^{5/3} (2700/\rho)/(2.4 \times 10^{-10} {\rm d}r/r \, {Q_D^*}^{5/6})$. $L_{\rm IR}/L_\star$ is the fractional luminosity of the debris disc, $e$ the mean 
eccentricity of the parent belt planetesimals, ${\rm d}r$ the belt width (in au), $\rho$ their bulk density (in kg/m$^3$) and $Q_D^*$ their collisional strength (in J/kg). For the purpose of this study, we use typical values \citep[as in][]{2008ARA&A..46..339W}, i.e we fix $e=0.05$, ${\rm d}r/r=0.5$, $\rho=3000$ kg/m$^3$, $Q_D^*=500$ J/kg, which
gives $\delta=2.9 \times 10^{5}$. We will 
study in subsection \ref{errorbarco} what change can result from varying these parameters. Therefore, the CO mass rate can be estimated as

\begin{equation}
\label{COloss}
\dot{M}_{\rm CO}=\gamma \dot{M}_{\rm loss},
\end{equation}

\noindent where $\gamma$ needs to be on the order of a few percent to fit the observed $\beta$ Pic CO mass or the composition of Solar System comets. More precisely, we fix $\gamma$ to 6\%, the upper limit found by \citet{2016MNRAS} for $\beta$ Pic (when taking into account that CO$_2$ dissociation can also contribute to observed CO). This is consistent
with the composition of Solar System comets for which $2\% < \gamma < 27\% $ \citep[when also including CO$_2$ that can contribute to the observed CO,][]{2011ARA&A..49..471M,2016MNRAS}.

To get the actual CO mass, one needs to know the photodissociation timescale $t_{\rm ph}$, which is directly proportional to the impinging UV radiation on the gas disc.
The main contributors to UV photons are the central star and the interstellar radiation field. The mean intensity field (in W/m$^2$/Hz) is defined as

\begin{equation}
\label{Jnu}
J_\nu = \frac{1}{4 \pi} \int_\Omega I_\nu \, {\rm d}\Omega,
\end{equation}

\noindent where the intensity $I_\nu$ is the sum of the stellar $I_\star$ and IRF $I_{\rm IRF}$ intensities, and is integrated over the solid angle $\Omega$ subtended by its source. We use \citet{2004astro.ph..5087C} stellar spectra for $I_\star$ and the Draine interstellar radiation field for $I_{\rm IRF}$ \citep{2011piim.book.....D}. 

Also, we take into account any attenuation of the flux coming from the star and IRF. When CO photodissociates, it creates atomic carbon and oxygen that spread all the way to the star. CI will photoionize by absorbing strong UV photons with energies greater than 11.26eV (the ionization potential of CI).
This will attenuate the UV flux impinging onto CO and reduce the CO photodissociation efficiency. We take into account the attenuation
in the radial direction for radiation coming from the star but also in the vertical direction for the IRF. We note that we do not attenuate the photons with energies lower than 11.26eV that may still participate in photodissociating CO. 
The new fluxes after attenuation are $I_\star \exp(-\tau_r)$ and $I_{\rm IRF} \exp(-\tau_v)$, where $\tau_r$ and $\tau_v$ are the radial and vertical optical thicknesses to UV radiation defined as

\begin{align}
\label{expInu}
  \tau_r(R)=\sigma_{\rm ion} \int_{0}^R  \, n_{\rm C_{I}}(R') \, dR',\\
  \tau_v(R)=\sigma_{\rm ion} \, H(R) \, n_{\rm C_{I}}(R),
\end{align}

\noindent where $n_{\rm C_{I}}$ is the CI number density, $\sigma_{\rm ion}$ is the CI ionization cross section and $H$ the height of the gas disc. Note that equations 4 and 5 require knowledge of the density of CI, the calculation of which is given in equations ~\ref{nc} and \ref{ionfrac} 
of section \ref{carbon}. For simplicity we first present here all the equations relating to CO, but note that the full model requires section \ref{carbon} to close the system of equations presented in this section. However, for most targets the CO photodissociation timescale will in fact be 
dominated by the interstellar radiation field (roughly outside of the green hatched area in Fig.~\ref{fig1}) and can be assumed to be 120yr so that Eqs. 4 and 5 are not needed to compute this timescale (see also KWC16).

The CO photodissociation timescale can now be computed

\begin{equation}
\label{phodi}
t_{\rm ph} = \left( \sum_{\nu=\nu_i} \frac{4 \pi J_\nu}{h \nu} \sigma_\nu^{\rm CO} \right)^{-1},
\end{equation}

\noindent where $\sigma_\nu^{\rm CO}$ is the CO photodissociation cross section per unit wavelength and $\nu_i$ are the frequencies of the lines that produce photodissociation (mostly in the UV). The cross sections are taken from \citet{2008CP....343..292V}.

\begin{figure}
   \centering
   \includegraphics[width=8.5cm]{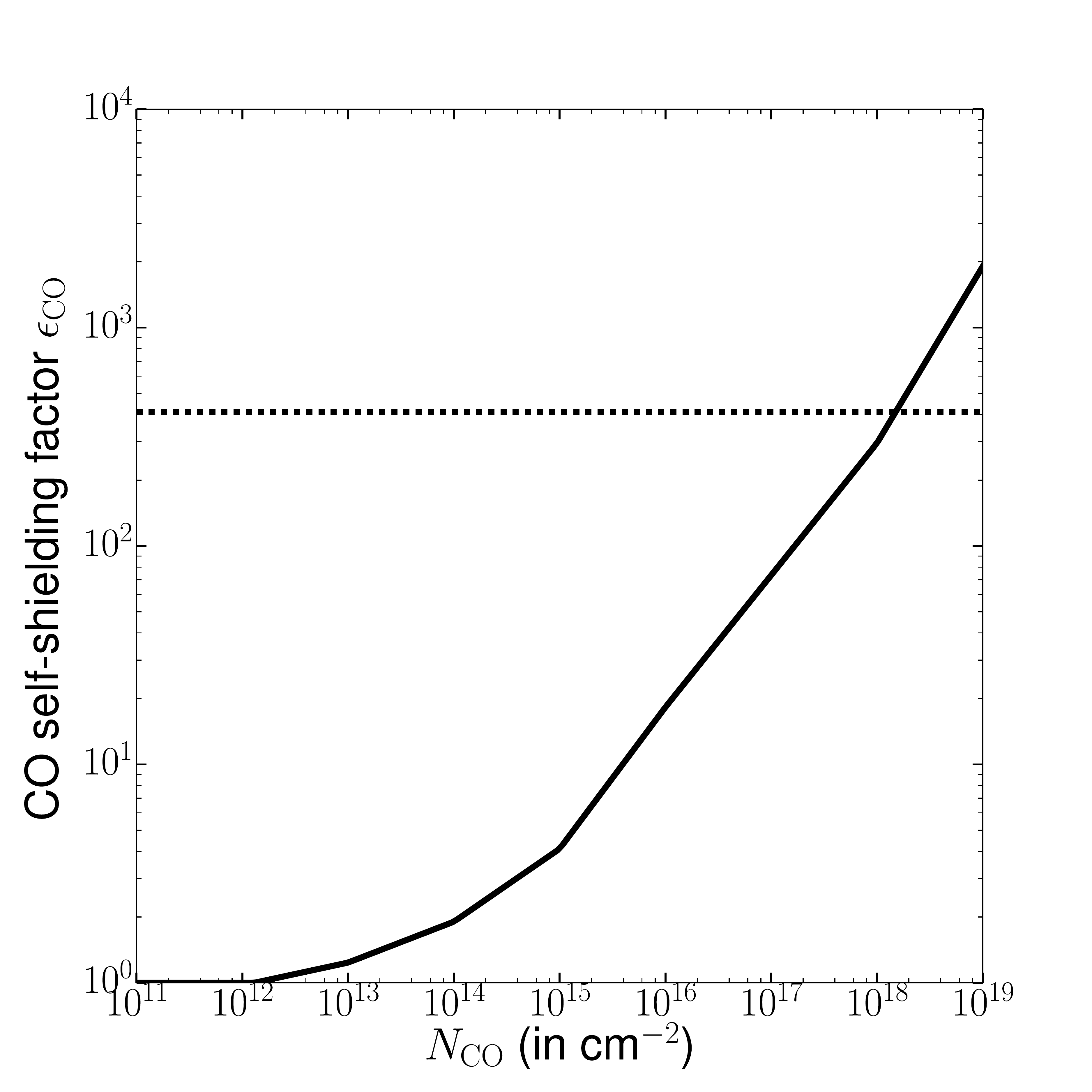}
   \caption{\label{figselfshi} CO self-shield factor $\epsilon_{\rm CO}$ Vs CO column densities $N_{\rm CO}$ (in cm$^{-2}$) in the vertical direction (solid line). Values taken from \citet{2009A&A...503..323V} and interpolated. The dashed line shows the maximum
$\epsilon_{\rm CO}$ that can be reached before photodissociation and viscous timescales are equal, limiting the growth of $\epsilon_{\rm CO}$ (see text for details).}
\end{figure}

Also, CO can self-shield against photodissociation if CO column densities in the vertical direction $N_{\rm CO}$ are $ \gtrsim 10^{12}$ cm$^{-2}$ \citep{2009A&A...503..323V}. We define the self-shielding
factor $\epsilon_{\rm CO}$ as being equal to one when CO is optically thin to UV radiation and scales as shown on Fig.~\ref{figselfshi}. For this calculation we assume that radiation is coming from all directions (from the interstellar radiation field), that there is no H$_2$ around (as it is secondary gas), that $T_{\rm ex}$=5K (small NLTE excitation temperature)
and that the CO linewidth is 0.3km/s. However, for a specific system, one can use different values for the linewidth or $T_{\rm ex}$ \citep[see Fig.~3 in][]{2009A&A...503..323V} to refine the estimate of $\epsilon_{\rm CO}$, which could vary by a factor $\sim$ 1.5.
The total mass of CO in the disc at any one time $M_{\rm CO}$ (in M$_\oplus$) is calculated assuming a steady state balance of gas production and loss so that

\begin{equation}
\label{mco}
M_{\rm CO}=\dot{M}_{\rm CO} t_{\rm ph} \epsilon_{\rm CO},
\end{equation}

\noindent where the $\epsilon_{\rm CO}$ factor accounts for the fact that the photodissociation timescale from Eq.~\ref{phodi} must be increased by this factor due to self-shielding.

\begin{figure}
   \centering
   \includegraphics[width=8.5cm]{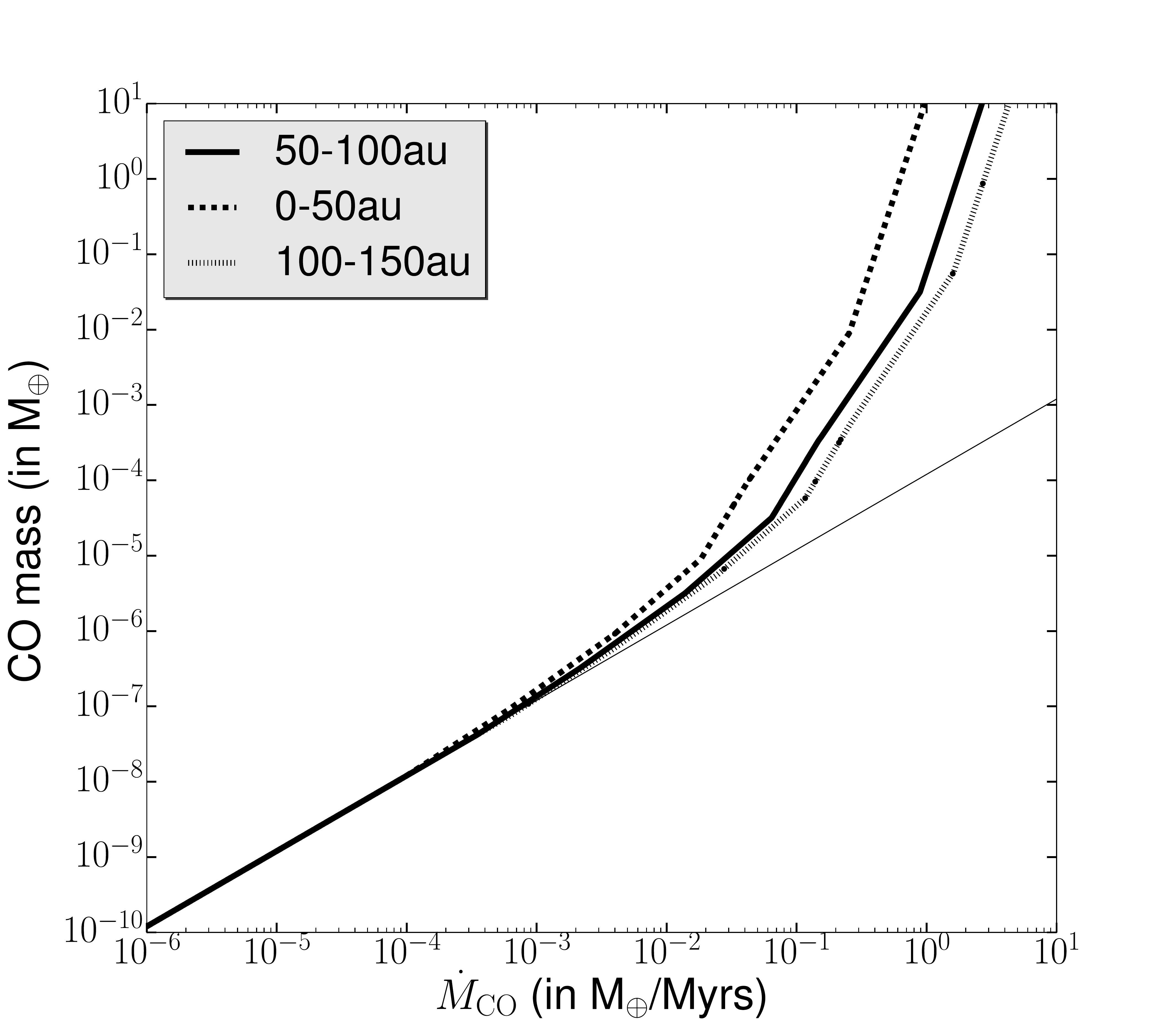}
   \caption{\label{figmcomdot} $M_{CO}$ (in M$_\oplus$) Vs $\dot{M}_{\rm CO}$ (in M$_\oplus$/Myrs) taking into account CO self-shielding at high input rates. We compute the relation for three disc locations, from 0-50au (dashed), from 50-100au (solid) and from 100-150au (dotted).
The thin line shows the unshielded values.}
\end{figure}

To check from which CO input rate self-shielding starts to matter, we plot $M_{\rm CO}$ as a function of $\dot{M}_{\rm CO}$ in Fig.~\ref{figmcomdot}. We assume that $t_{\rm ph}=120$ years and use 
different disc locations (from 0-50au in dashed, from 50-100au in solid and from 100-150au in dotted) to convert from $M_{\rm CO}$ to $N_{\rm CO}$ (CO column density in the vertical direction) assuming a constant surface density in the disc. We iterate a couple of times as $M_{\rm CO}$ depends on $N_{\rm CO}$, to
reach convergence. For $\dot{M}_{\rm CO} \gtrsim 5 \times 10^{-3}$ M$_\oplus$/Myrs, this effect will become important and the CO mass will increase steeply.

However, when CO self-shielding is important, CO photodissociation timescales become very long and CO may have time to spread viscously, hence reducing the vertical column density. We also  implement self-shielding into our model. 
To do so, we compute the viscous timescale (see Eq.~\ref{viscous}, where we assume $\alpha=0.5$) for each system (depending on the location of the parent belt) and compare it to the CO photodissociation timescale that includes self-shielding. If the latter becomes longer than the viscous timescale, we assume
no more shielding from CO and keep the value where these two timescales are equal. We added a dashed line in Fig.~\ref{figselfshi} showing the maximum $\epsilon_{\rm CO}$ that can be reached before photodissociation and viscous timescales are equal assuming $R_0=85$au, $T$=100K 
and $\alpha=0.5$ around a $\beta$ Pic-like star. Therefore, the CO self-shielding factor $\epsilon_{\rm CO}$ cannot grow to extremely large values.

\begin{figure*}
   \centering
   \includegraphics[width=16.5cm]{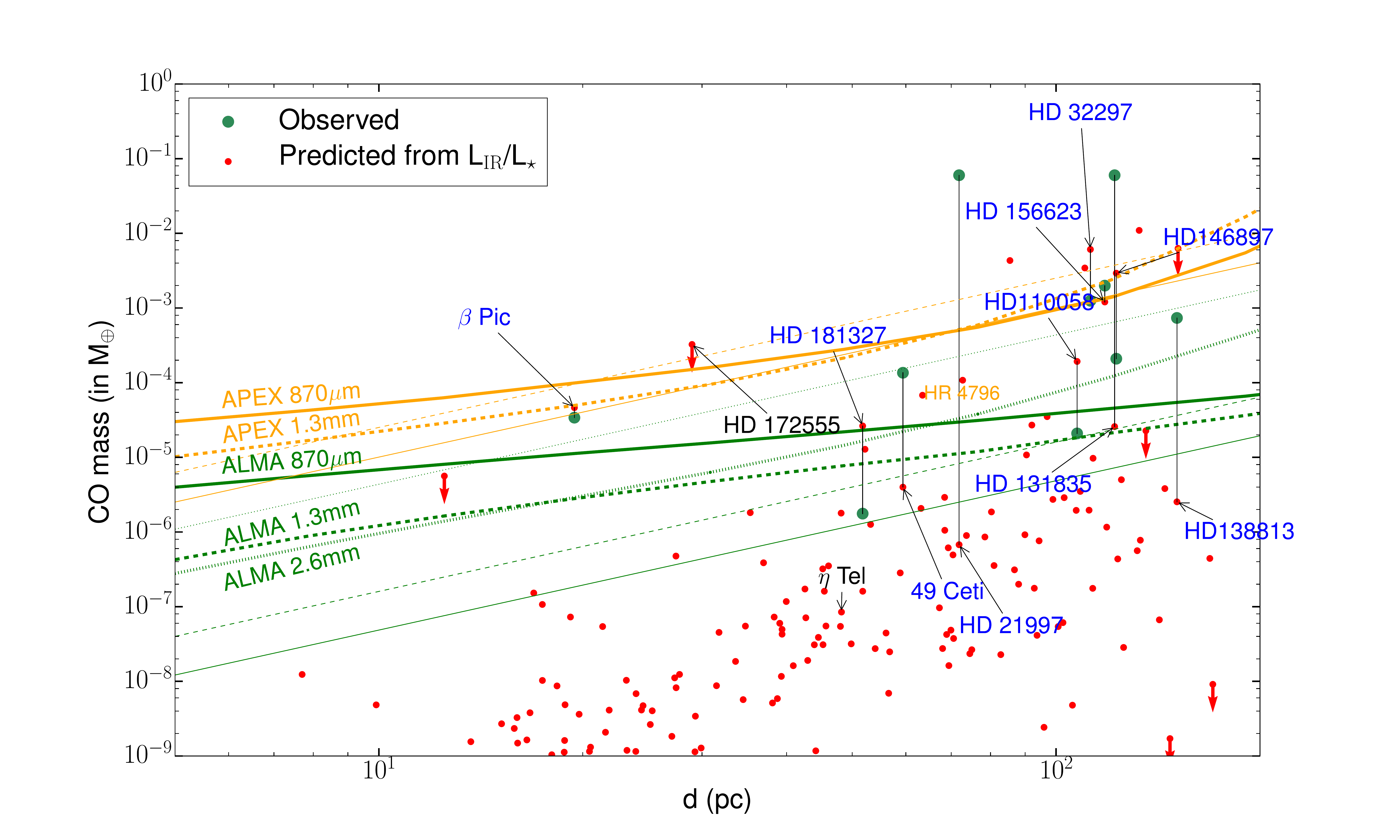}
   \caption{\label{figCO} CO mass (in M$_\oplus$) as a function of distance to Earth ($d$). Planetary systems with gas detections are labelled with their names. If CO is detected, the label is in blue (black otherwise). 
The CO mass worked out from observations are shown as green points (see Table~\ref{tab4}).
The red points are predictions from our model. The red downward arrows show systems that are in the blue hatched area on Fig.~\ref{fig1}, which cannot keep CO trapped on solid bodies. The thin orange lines show detection limits (assuming LTE and $T$=100K) with APEX at 1.3mm and 870$\mu$m (5$\sigma$ in one hour) and the thin green lines
are for ALMA at the same wavelengths and 2.6mm. We also compute the detection thresholds in NLTE, assuming an ionization fraction of 0.1 and $R_0$=85au. The corresponding thick lines are in orange for APEX and green for ALMA and we keep the same line style for the different wavelengths.}
\end{figure*}

\subsection{CO predictions compared with observations}\label{COcomp}

We compare our CO predictions (small red points using Eq.~\ref{mco}) with observed CO masses (large green points) in Fig.~\ref{figCO} that are computed from the most recent CO integrated line fluxes in NLTE (see Table \ref{tab1} for references). 
On this plot, the systems with CO detected are also labelled in blue. Other systems with gas detected but no CO are labelled in black. HR 4796 is labelled in orange
for informational purposes, as it stands out in most of our predictions (except for CO), but no gas has been detected yet. Observed masses and predictions are linked by a thin black line for each individual system with CO detected. We also plot detection thresholds for APEX (orange) and ALMA 
(green) at different wavelengths to check for the detectability of each individual system but we detail that in a coming subsection \ref{detect}. For now, we focus on comparing mass predictions with observed masses.

Comparing the CO predictions with observations, we find that 7/10 systems have $0.1 \lesssim M_{\rm observed}/M_{predicted} \lesssim 10$, and can be explained by a secondary gas model. Specifically, we find that $\beta$ Pic, HD 181327, 49 Ceti, HD 32297, HD 110058, HD 156623 and HD 146897 can be well explained with secondary gas being
produced within the debris belt already known to be present.

The 3/10 systems left have $M_{\rm observed}/M_{predicted} > 100$, while others are generally within a factor 10 or slightly more. We check later (see subsection \ref{errorbarco}) that varying the parameters of the model can account for a factor $\sim$ 10 difference but
we consider systems that have $M_{\rm observed}/M_{predicted} > 100$ as not possible to explain with a secondary gas model. Specifically, HD 21997, HD 131835 and HD 138813 are in this category. We note for instance that for HD 21997, for which the CO mass is relatively well-known 
(thanks to the detection of an optically thin C$^{18}$O line), the five orders of magnitude
that separate the observation from our prediction will never be accounted for by our model. This reinforces the conclusion of \citet{2013ApJ...776...77K}, who suggested that CO observations in this system can only be explained if the CO is of primordial origin. 
Thus, our model can also be used to identify systems that show anomalous behaviour when predictions do not match observations by orders of magnitude. For these systems, it may well indicate that they are of primordial origin.

Our model also finds that the predicted CO mass around $\eta$ Tel is well below the predictions for systems where CO gas is detected. This is owing to the early stellar type (A0V) of $\eta$ Tel, which reduces the CO photodissociation timescale to a small value so that CO cannot accumulate in this system. This
may explain the non-detection of CO in this system so far (see subsection \ref{detecta}).

We also find that the CO mass in HD 172555 is predicted to be relatively high ($\sim 3\times 10^{-4}$ M$_\oplus$) possibly at a detectable level. However, as the debris belt in this system is very close-in, it might be optically thick and the line flux not detectable with APEX 
(see subsection \ref{detecta}). Also, as explained in subsection \ref{firstc}, it may be that grains in this system are just too warm (in the blue hatched area in Fig.~\ref{fig1}) to retain any CO.

Our simple analytical model is thus able to reproduce CO observations (within uncertainties, see section \ref{errorbarco}) for these specific systems and to flag anomalous systems which may be of primordial origin or have a secondary origin that fails to be modelled by the gas production mechanism
assumed in this paper.

The variations between our predictions and observations were expected as we do not fit our model to the observations but rather use fiducial values to see 
if the model reproduces the bulk of observations. For instance, we assumed that $\gamma=6\%$ of dust is converted into CO \citep[as found for $\beta$ Pic][]{2016MNRAS}. This value might vary from one system to another and could 
potentially explain differences between some observed values and our predictions. For example, $\gamma$ may vary with age as the more CO is depleted from grains, the less is exposed and the less comes off grains. $\gamma$ also varies with initial composition, which depends on initial 
abundances in the extra-solar nebula in which grains formed, and depends also on planetesimal formation mechanisms. We note that gas detections provide a way to get back to the value of $\gamma$ and then to the amount of CO in planetesimals.

\subsection{Model predictions for a large sample of debris disc stars}

We selected a sample of 189 debris disc host stars to determine which of these is predicted to have CO at a detectable level according to our model. 
The goal of our star selection process is to choose any star for which gas is likely to be detectable. So the nearby stars were included as they are close, and the bright systems because they are still potentially 
detectable (as $\dot{M}_{\rm CO} \propto (L_{\rm IR}/L_\star)^2$) despite being farther away.
We assume the same fiducial parameters as in subsection \ref{COcomp} to compute CO masses. These systems without gas detected are also shown as red points in Fig.~\ref{figCO} but they are not labelled (except for HR 4796). If the systems lie in the blue hatched area in Fig.~\ref{fig1} 
(i.e the black body temperature of grains is greater than 140K), we use red downward arrows as CO masses predicted are then only upper limits. In these cases, we do not necessarily expect to be able to detect CO. 

The parameters assumed for the sample of stars can be found in Appendix \ref{tables}. Most debris disc systems in our sample are not spatially resolved and we cannot measure $R_0$ (the planetesimal belt location) directly from images.
Rather, for this sample, $R_0$ comes from an SED fit of each individual spectrum. For the parent belt radius, we do not use the black body radius, which is always 
underestimated but rather use Eq.~8 in \citet{2015MNRAS.454.3207P} to correct this radius (assuming a composition of 50\% ice and 50\% astrosilicate, see their table~4).

We see that the predicted CO masses for $\sim$ 20 stars in this sample are comparable to the level of observed CO masses in other systems. Other stars from the sample have lower CO masses that according to our model should not be detectable with current instruments. We will study in more
detail their detectability in the next subsection.

\begin{table}
    \caption{NLTE calculations of CO masses from observations (optically thin assumption). We indicate as a footnote where the line fluxes were taken from. The second column provides the assumed electron density (computed from the model presented in this paper) when computing the CO mass.}
\begin{center}

\begin{threeparttable}
\begin{tabular}{|l|c|c|}
  \toprule
  Star's  & CO mass  & Electron Density\\
  name & (M$_\oplus$) & (cm$^{-3}$)\\
  \midrule
$\beta$ Pic$^1$  & $2.8\times10^{-5}$& 240 \\
49 Ceti$^2$  & $1.4\times10^{-4}$& 350 \\
HD 21997$^3$  & $6.0\times10^{-2}$& 510 \\
HD 32297$^4$  & $1.3\times10^{-3}$& 360 \\
HD 110058$^5$ &  $2.1\times10^{-5}$& 500 \\
HD 131835$^6$ &  $6.0\times10^{-2}$& 10 \\
HD 138813$^5$  & $7.4\times10^{-4}$& 1400 \\
HD 146897$^5$  & $2.1\times10^{-4}$& 5 \\
HD 156623$^5$  & $2.0\times10^{-3}$& 10 \\
HD 181327$^7$  & $1.8\times10^{-6}$& 130 \\


  \bottomrule
\label{tab4}
\end{tabular}
\begin{tablenotes}
		\footnotesize
   		\item $^1$ \citet{2016MNRAS}, $^2$ \citet{2008ApJ...681..626H}, $^3$ \citet{2013ApJ...776...77K}, $^4$ \citet{2016MNRAS.461.3910G}, $^5$ \citet{2016ApJ...828...25L}, $^6$ \url{http://www.eso.org/sci/meetings/2016/Planet-Formation2016/Contributions/Oral/planets2016-MoorA.pdf}, $^7$ \citet{2016MNRAS.460.2933M}
  	\end{tablenotes}
\end{threeparttable}

\end{center}

\end{table}

\subsection{Detection thresholds for APEX and ALMA}\label{detect}

In this subsection, we first describe in \ref{dtc} how we computed the APEX/ALMA detection thresholds in Fig.~\ref{figCO}. In \ref{fluxpr}, we then explain how we compute flux predictions from our mass predictions. We can then assess in \ref{detecta} the detectability of the CO mass
predicted for each system. In \ref{vLTE}, we quantify the location of the LTE/NLTE transition. In \ref{vnLTE}, we explain in more detail the NLTE calculations used throughout the paper and describe how the NLTE regime varies compared to LTE. Finally,
in \ref{thick}, we explain how we compute the optical thickness of CO transitions.

\subsubsection{Detection threshold calculation}\label{dtc}
To assess the detectability of each system in Fig.~\ref{figCO}, 
we compute the detection thresholds with APEX (orange lines) and ALMA (green lines), assuming gas both in LTE (thin lines, see subsection \ref{vLTE}) and NLTE (thick lines, see subsection \ref{vnLTE}). While NLTE provides
the most accurate estimate of the line fluxes, LTE might be a valid approximation in certain regions of parameter space and is much simpler to calculate. Thus we plot both to emphasise their differences \citep[see][]{2015MNRAS.447.3936M}. 

For an optically thin line, the integrated line flux seen at Earth is

\begin{equation}
\label{lineflux}
F_{u,l}=\frac{h \nu_{u,l} A_{u,l} x_u M}{4 \pi d^2 m},
\end{equation}

\noindent where $A_{u,l}$ is the Einstein A coefficient for spontaneous emission, $\nu_{u,l}$ the frequency of the transition and $x_u$ the fraction of molecules that are in the upper energy level $u$, $M$ the total mass and m the mass of the studied molecule (or atom). We assume a typical gas temperature of 100K and show assumed sensitivities ($5\sigma$ in one hour from
the APEX and ALMA online calculators) in Table~\ref{tabsens}. The assumed PWV and elevation are given in the description of Table~\ref{tabsens}. The sensitivity is, here, independent of baseline configuration as we assumed that the gas discs would be unresolved, 
which yields maximum detectability.
Using a Boltzmann distribution for the LTE case (thin orange and green lines) or solving the full statistical equilibrium (NLTE, thick orange and green lines), one can find the total number of CO molecules (or CO mass) that would create a certain flux at Earth. 

If the LTE approximation is valid, Fig.~\ref{figCO} shows that
for a gas temperature of 100K, the 870 $\mu$m transition is more sensitive than the 1.3mm and 2.6mm transitions (but this changes in NLTE) and the ALMA detection threshold is better than APEX by two orders of magnitude. However, 
for most debris disc systems, CO is likely out of LTE \citep{2015MNRAS.447.3936M}. In this case, detection limits depend on the electron density (which can be computed from our model, see section \ref{carbon}) and do not simply scale as $d^2$ (see thick lines). Also, for the NLTE detection thresholds, 
we take account of the optical thickness of lines as explained in subsection \ref{thick}. 

\subsubsection{Flux predictions}\label{fluxpr}
In Table~\ref{tab2}, we provide mass and flux predictions for the targets with the largest predicted CO fluxes (and later in Appendix \ref{tables} for all our targets). We use the CO NLTE code developed in \citet{2015MNRAS.447.3936M} to convert from a given CO mass (outcome of our model)
to a flux observed from Earth. To do so, we must provide to the code the amount of radiation seen by the gas. When modelling low rotational transitions of CO, their excitation is likely to be dominated by the CMB \citep[see][]{2015MNRAS.447.3936M} and our conversion will be accurate. 
We also account for the optical thickness of lines along the line-of-sight to Earth in our flux calculations using the method described in subsection \ref{thick}. 

When making predictions for CI, CII and OI masses, the conversion to fluxes is more complicated (see subsection \ref{fluxat}).

\subsubsection{Detectability of the specific sources used in this paper}\label{detecta}
We can now look at the relative position of our CO mass predictions against the detection thresholds in Fig.~\ref{figCO}. Red points need to be above the ALMA NLTE detection threshold for a given transition to be detected at $>$5$\sigma$ within one hour. We note that the NLTE lines assume
a certain electron density (computed assuming a carbon ionization fraction equal to 0.1 and that $R_0$=85au, KWC16) so these lines can move up or down if the electron density is smaller or higher, respectively (unless the lines are in a purely radiative regime where only the CMB excites the lines). 
That is why we leave the LTE detection thresholds (best case scenario) as a guide to check the range of detection thresholds that could be spanned if more colliders were around. 

We find that systems that are predicted to be detectable by APEX are also close to being optically thick. Both the low sensitivity and high optical thickness explains the difficulty to detect CO with single-dish aperture telescopes.

Our model explains why $\beta$ Pic was not detected by single dish telescopes \citep{1995MNRAS.277L..25D,1998A&A...334..935L} as it lies below the APEX detection limits. CO was detected with APEX around HD 21997 \citep{2011ApJ...740L...7M} and HD 131835 \citep{2015ApJ...814...42M}
and with the JCMT around HD 32297 \citep{2016MNRAS.461.3910G}. We predict that it should be the case for HD 32297. However, HD 21997 and HD 131835 are not predicted to be detectable and yet CO was detected. 
This is because they have larger masses than predicted likely because the CO is primordial. 

We can now check whether the CO APEX non-detection in $\eta$ Tel \citep{2015ApJ...814...42M} is predicted by our model. Our prediction lies well below the APEX detection threshold. Even ALMA should not be able to detect CO as
there is roughly two orders of magnitude between the 1.3mm NLTE limit and the CO mass inferred by our model.

Looking at Table~\ref{tab2}, we see that if CO can remain on grains, HD 172555 is one of the most promising targets to look for CO. Though, we note that our flux prediction is still lower (by a factor 5) than the APEX upper limit of $1.6 \times 10^{-20}$ W/m$^2$ 
\citep{2011ApJ...740L...7M}. This is because the CO in this system is very close in and so optically thick (see subsection \ref{thick} for more details on how the optical thickness was computed).

From our mass predictions, $\sim$3 more systems (HD 114082, HD 117214, and HD 129590) are above the APEX detection threshold plotted in Fig.~\ref{figCO}. However, that detection threshold was computed for a system at 85au with an ionization fraction of 0.1, and
computing the fluxes of these 3 systems at the correct radius, we find that their CO is optically thick and so not detectable with APEX. 
These stars are part of the Sco-Cen association and were observed recently with ALMA, but this led to no detections \citep{2016ApJ...828...25L}. These CO observations reached a sensitivity of $\sim 5 \times 10^{-22}$ W/m$^2$ (5$\sigma$) for these 3 systems
which is still a factor 2 below our predictions. We note that for these 3 systems, CO self-shielding is high but limited by viscous spreading of CO.
  
The NLTE detection thresholds for ALMA are at least 10 times more sensitive in mass than for APEX at a given distance $d$. This is not only due to the different instrument sensitivities but also to the electron density and $\tau_\nu$ being smaller for systems that lie close to the ALMA detection thresholds
compared to systems that are close to the APEX thresholds.
 Our model predicts that $\sim$ 15 systems from the sample lie above or close to the ALMA detection
thresholds. Furthermore, the NLTE lines could be closer to the LTE regime for systems that are closer-in than $R_0$=85au, due to the higher electron density. Therefore, systems under the NLTE lines could still 
be detectable. We provide a list of the 15 most promising systems in Table~\ref{tab2} for which we predict CO could be detected with ALMA. For instance,
we see that the CO around HR 4796A (labelled in orange) as well as HD 15745 may be detectable with ALMA.

\subsubsection{Validity of LTE}\label{vLTE}
To understand when CO is out of LTE, we computed the CO-electron critical density (for the different transitions) for an optically thin system assuming a two level system. This critical density is simply equal to $A$/$\gamma_c$, where $A$ is the Einstein
coefficient of the considered transition and $\gamma_c$ the collisional rate coefficient \citep[from upper to lower level, taken from][]{1975JPhB....8.2846D}. We can compute the CO mass $M_{\rm CO,LTE}$ required by our model to create
an electron density equal to this critical density. 
To compute the LTE limit, we further assume that all electrons come from CI photoionization, that the ionization fraction equals 0.1, that $R_0=85$au and that the photodissociation timescale is $\sim 120$ years (and then use Eq.~\ref{nc} derived later). 

We find that the LTE limit (for the 870$\mu$m transition, solid line) is located at $M_{\rm CO} \sim 5 \times 10^{-4}$ M$_\oplus$ in Fig.~\ref{figCO}, above which LTE likely 
applies (i.e. in almost no systems).  Note that $M_{\rm CO,LTE}$ scales as $T_{\rm gas}^{3/2} t_{\rm ph} f^{-1} L_\star^{-1/4} R_0^3$ and so the LTE limit is different
for different systems. For instance, for a system with a debris belt at $R_0=10$au, the LTE limit might be expected to go down by a factor 600. However, for such close-in systems, the ionization fraction would also drop (because of higher CI densities closer in). 
So overall, our prediction that almost all debris discs with gas are not in LTE for CO is generally true, reinforcing the conclusion of \citet{2015MNRAS.447.3936M}, and motivating the need for
line ratios to test if the gas has an exocometary origin \citep{2016MNRAS}.

\subsubsection{Calculations in non-LTE}\label{vnLTE}
We here give more details on how we computed the NLTE detection thresholds but also give the reader a feel for the differences it implies compared to the LTE regime. We used the code
presented in \citet{2015MNRAS.447.3936M} to solve the statistical equilibrium and work out the population of rotational levels. For the low CO transitions considered here, it was shown in \citet{2015MNRAS.447.3936M} that for $\beta$ Pic, the excitation will be dominated by the CMB radiation rather than dust emission and stellar 
radiation (see Fig.~\ref{figrad}). The same also applies to other systems as $\beta$ Pic is among the most luminous debris discs and the CMB is even more dominant in less dusty systems.

\begin{figure}
   \centering
   \includegraphics[width=8.5cm]{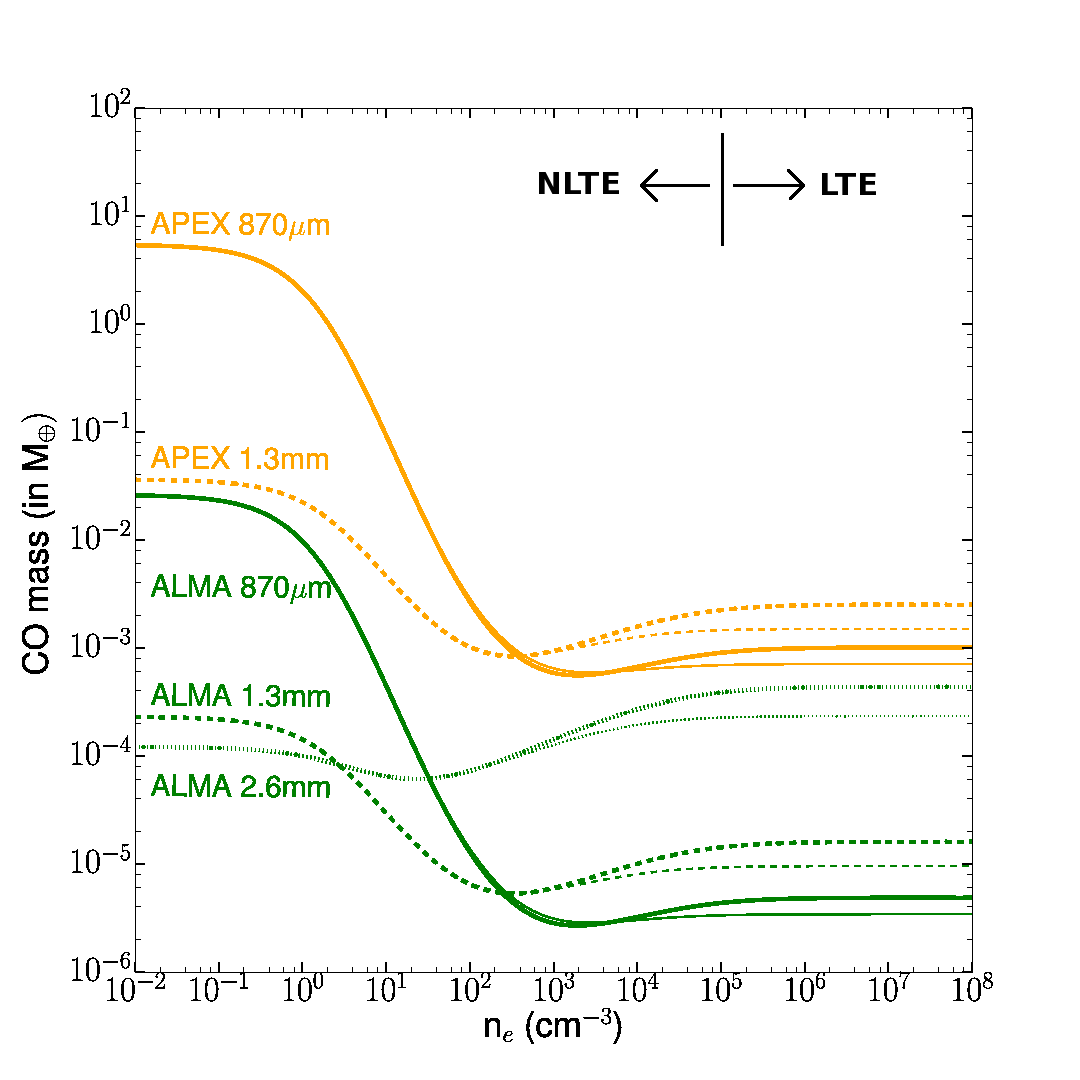}
   \caption{\label{figmcoel} Mass in CO (in M$_\oplus$) required to result in a predicted line flux equal to the sensitivity threshold of APEX (in orange) and ALMA (in green) as a function of electron density (in cm$^{-3}$). The assumed sensitivities 
for different transitions that can be found in Table~\ref{tabsens}. Thinner lines are for $T$=50K and thicker lines for $T$=100K. The black vertical line shows the transition between the LTE and NLTE regimes.}
\end{figure}

Looking at the differences with LTE is instructive.
The APEX sensitivity lines are close to LTE at larger distances. However, there are not enough colliders (assumed to be electrons) to be in full LTE even though the lines are above the LTE limit derived above (located at $M_{\rm CO} \sim 5 \times 10^{-4}$ M$_\oplus$). This is owing to the two population level assumption made
when computing the LTE limit. Also, for almost all distances, the APEX 1.3mm transition in NLTE is actually more sensitive than LTE. This is expected as for these given higher masses or electron densities ($10^1-10^4$ cm$^{-3}$), the CMB excites the 1.3mm transition more than collisions do (see Fig.~\ref{figmcoel}).
Also, the 2.6mm transition is much more excited in NLTE for electron densities smaller than $\sim 10^5$ cm$^{-3}$ (see Fig.~\ref{figmcoel}). This was already shown in \citet{2015MNRAS.447.3936M} in their Fig. 6. Therefore, quite strikingly, for very low electron densities, ALMA is more sensitive in the 2.6mm transition rather than the 870$\mu$m. 

We also notice that in NLTE, the 1.3mm transition is more sensitive than the 870$\mu$m transition for lower masses (unlike the LTE case). This is also expected from Fig.~\ref{figmcoel}, which shows the mass needed to reproduce a given flux as a function of the electron density. Indeed, for electron densities $\lesssim 5\times10^{2}$ cm$^{-3}$, the
1.3mm transition needs less CO mass than the 870$\mu$m transition and is thus more sensitive. Hence, when observing a particular system, one should consider carefully which transition is more suited.

\begin{table}
    \caption{Instrument sensitivities to reach 5$\sigma$ in one hour on source (overhead excluded) at 45$^{\circ}$ elevation assuming unresolved gas discs. We assumed a PWV of 2mm for CO observations and 0.5mm for CI.
We used 40 antennas for ALMA. For species observed with Herschel,
we assumed the typical sensitivities reached by real observations \citep{2014A&A...565A..68R}. For SPICA and for FIRS the sensitivities are taken from the SPICA and FIR surveyor documentations.}
\begin{center}

\begin{threeparttable}
\begin{tabular}{|l|l|c|}
  \toprule
  Instrument & Line & Sensitivity (W/m$^2$) \\
  \midrule

APEX & CO (1.3mm) & $3.3\times10^{-20}$\\
 & CO (870$\mu$m) &  $8.5\times10^{-20}$\\
 & CI (610$\mu$m) &  $3.9\times10^{-19}$\\
 & CI (370$\mu$m) &  $1.2\times10^{-18}$\\
ALMA & CO (2.6mm) &  $2.0\times10^{-22}$\\
 & CO (1.3mm) &  $2.1\times10^{-22}$\\
 & CO (870$\mu$m)  &  $4.1\times10^{-22}$\\
 & CI (610$\mu$m) &  $2.5\times10^{-21}$\\
 & CI (370$\mu$m) &  $7.9\times10^{-21}$\\

Herschel/PACS & CII (158$\mu$m) & $8\times10^{-18}$\\
 & OI (63$\mu$m) &  $6\times10^{-18}$\\
SPICA/SAFARI & CII (158$\mu$m) & $3\times10^{-19}$\\
 & OI (63$\mu$m) &  $3\times10^{-19}$\\
FIRS (10m) & CII (158$\mu$m) & $1.5\times10^{-21}$\\
 & OI (63$\mu$m) &  $1.5\times10^{-21}$\\

  \bottomrule
\label{tabsens}
\end{tabular}
\end{threeparttable}

\end{center}

\end{table}

\subsubsection{Optical thickness of lines}\label{thick}

When considering detectability, one has to take into account the optical thickness of CO lines. An optically thick line will saturate and become harder to detect given its mass compared to an optically thin line. To estimate the optical thickness $\tau_\nu$ for a given CO mass, we 
assume a rectangular line profile and use the $\tau_\nu$ definition \citep{2016MNRAS} 

\begin{equation}
\label{tau}
\tau_\nu=\frac{h \nu_{ul}}{4 \pi \Delta \nu} (x_l B_{lu} - x_u B_{ul}) N,
\end{equation}

\noindent where $\Delta \nu$ is the linewidth in Hz, $x_u$ and $x_l$ are the fractional populations of the upper and lower levels of the given transition, $B_{lu}$ and $B_{ul}$ are the Einstein $B$ coefficients for the upward and downward transitions 
(that can be expressed as a function of the Einstein $A$ coefficient) and $N$ is the column density along the line-of-sight. We fix $\tau_\nu=1$ and compute the column density $N$ of CO needed to become optically thick. In order to do this, we assume a disc located between 70 and 100au, 
with a constant surface density, and a constant scale height equal to 0.2$R$ \citep[as found for $\beta$ Pic,][]{2012A&A...544A.134N} and work out the CO mass needed to reproduce this column density along the densest line-of-sight for an edge-on configuration. 
We choose the linewidth to be 2 km/s, which is close to the intrinsic 
linewidth found for $\beta$ Pic \citep[see][]{1994MNRAS.266L..65C,2014A&A...563A..66C}. This linewidth is the combination of thermal and turbulent broadening. This $\tau_\nu=1$ line
will vary depending on the extension of the disc, its scale height, the linewidth and gas temperature. Also, we assumed LTE to compute the population levels. We can check a posteriori that the approximation works as we find that the $\tau_\nu=1$ line for CO 
masses is $\sim 8 \times 10^{-4}$M$_\oplus$, which is dense enough to be above the LTE threshold (see subsection \ref{vLTE}). 

Thus, we are able to compute $\tau_\nu$ for every given mass and transition in Fig.~\ref{figCO} and take the optical thickness into account when computing mass detection limits from the telescope flux sensitivities. We applied the correction when plotting the NLTE detection thresholds in Fig.~\ref{figCO},
assuming that systems are edge-on, by applying a factor $\tau_\nu/(1-e^{-\tau_\nu})$ to the previously computed mass detection limit. Therefore, when $M_{\rm CO}$ is much above the $\tau_\nu=1$ limit, the mass detection limit at a given distance is increased by a factor $\tau_\nu$. One can see that our NLTE
detection limits in Fig.~\ref{figCO} start increasing more steeply with distance when approaching CO masses of $10^{-3}$M$_\oplus$ due to this reason, which hinders CO detections with APEX for targets at large distances. $\tau_\nu$ is computed for every transition and the corrections are applied with the corresponding $\tau_\nu$. This is why the APEX 1.3mm
line starts steepening before the 870$\mu$m line.

\begin{figure*}
   \centering
   \includegraphics[width=18.5cm]{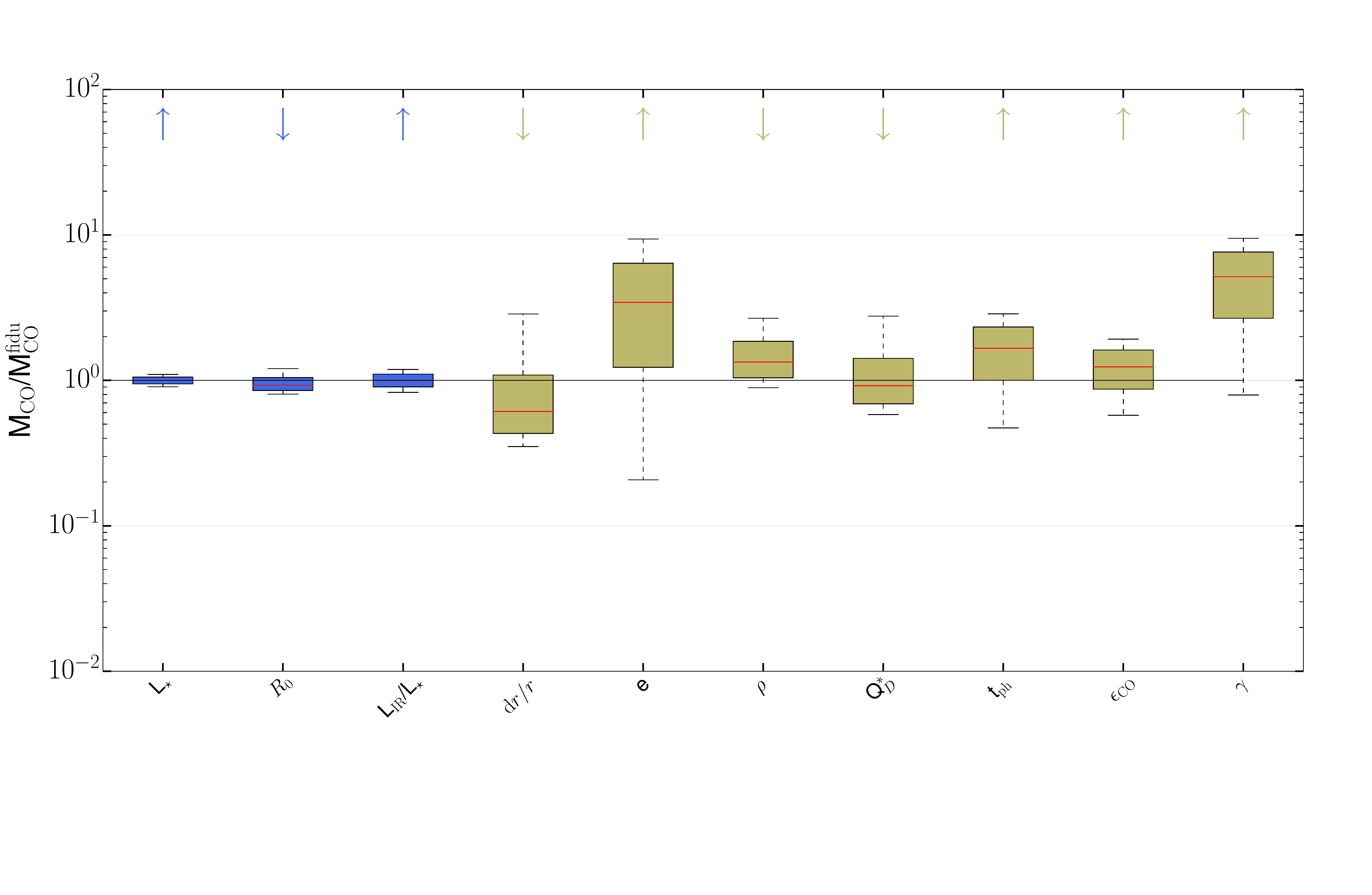}
   \caption{\label{figvarco} CO mass variation whilst varying parameters one by one. The fiducial values and amplitude of variations of the different parameters are given in Table~\ref{taberrorbar}. The blue boxes are for parameters where we have a good observational handle and only error bars on the predicted 
values are taken into account. For fawn boxes, there are more uncertainties and we allow for a large physical variation for each parameter. Box sizes show where 50\% of the distribution is located, while the whiskers contain 95\% of the distribution. The red line is the median value. We indicate with an upwards or
downwards arrow the sense of variation for $M_{\rm CO}$ if a given parameter is increased.}
\end{figure*}

\begin{table}
    \caption{List of ALMA promising targets to look for CO and their predicted masses and fluxes at 1.3mm and 870$\mu$m.}
\begin{center}

\begin{threeparttable}
\begin{tabular}{|l|c|c|c|}
  \toprule
  Star's  & CO mass & $F_{\rm CO}$ \tiny{1.3mm} &  $F_{\rm CO}$ \tiny{870$\mu$m} \\
  name & (M$_\oplus$) & (W/m$^2$) & (W/m$^2$) \\
  \midrule

HR 4796 & 1.1$\times 10^{-04}$ & 3.8$\times 10^{-21}$ & 1.4$\times 10^{-20}$ \\
HD 15745 & 6.8$\times 10^{-05}$ & 2.1$\times 10^{-21}$ & 2.5$\times 10^{-21}$ \\
HD 172555 & 3.2$\times 10^{-04}$ & 1.9$\times 10^{-21}$ & 3.2$\times 10^{-21}$ \\
HD 114082 & 4.3$\times 10^{-03}$ & 1.2$\times 10^{-21}$ & 1.9$\times 10^{-21}$ \\
HD 191089 & 1.3$\times 10^{-05}$ & 1.1$\times 10^{-21}$ & 1.1$\times 10^{-21}$ \\
HD 129590 & 1.1$\times 10^{-02}$ & 9.9$\times 10^{-22}$ & 1.6$\times 10^{-21}$ \\
HD 117214 & 3.4$\times 10^{-03}$ & 9.2$\times 10^{-22}$ & 1.6$\times 10^{-21}$ \\
HD 106906 & 2.7$\times 10^{-05}$ & 6.9$\times 10^{-22}$ & 7.8$\times 10^{-22}$ \\
HD 69830 & 5.6$\times 10^{-06}$ & 4.6$\times 10^{-22}$ & 9.2$\times 10^{-22}$ \\
HD 121191 & 6.3$\times 10^{-03}$ & 3.3$\times 10^{-22}$ & 8.7$\times 10^{-22}$ \\
HD 95086 & 1.1$\times 10^{-05}$ & 2.2$\times 10^{-22}$ & 8.4$\times 10^{-23}$ \\
HD 143675 & 9.7$\times 10^{-06}$ & 1.5$\times 10^{-22}$ & 2.3$\times 10^{-22}$ \\
HD 61005 & 1.8$\times 10^{-06}$ & 1.4$\times 10^{-22}$ & 2.2$\times 10^{-23}$ \\
HD 169666 & 1.3$\times 10^{-06}$ & 1.3$\times 10^{-22}$ & 2.9$\times 10^{-22}$ \\
HD 221853 & 2.9$\times 10^{-06}$ & 1.2$\times 10^{-22}$ & 5.0$\times 10^{-23}$ \\

  \bottomrule
\label{tab2}
\end{tabular}
\end{threeparttable}

\end{center}

\end{table}

\subsection{CO mass variation when changing parameters}\label{errorbarco}
Fig.~\ref{figvarco} can be used to work out the effect of varying one parameter of the model while keeping others fixed. For each parameter, the fiducial values and range of variations used to make the plot are listed in Table~\ref{taberrorbar}. The downwards and upwards arrows in Fig.~\ref{figvarco}
show the sense of CO mass variation if a parameter is increased. The blue boxes are for parameters that can be deduced from observations. For those, the rate of variation will correspond to the error bars from observations. We assume that L$_\star$ is known within 10\% \citep{2015A&A...582A..49H}, $R_0$
is computed from the SED (temperature) for the sample, and is known within a factor 2 \citep{2015MNRAS.454.3207P}. Also, if the SED has more than 2-3 far-IR detections (which is the case for the sample we use), the fractional luminosity is known within about 10\% 
(most discs in our sample are bright so have far-IR photometry with a signal-to-noise ratio $>$10, meaning that the disc temperature and normalisation, and thus the fractional luminosity, are well constrained). 

For the fawn boxes in Fig.~\ref{figvarco}, 
we vary the parameters over a larger region (see Table~\ref{taberrorbar}). Note that each parameter is varied across a reasonable range for the given parameter so that the planetesimal eccentricity varies by a factor 20, while $L_\star$ varies by 10\% and $t_{\rm ph}$ by 300\% (because
of uncertainties on the interstellar radiation field around these far-away discs). 
We assume a uniform distribution while varying each parameter. We then compute the result for each variation. Then, the box sizes show where 50\% of the distribution lies and the whiskers contain 95\% of the distribution. The red line shows the median of the distribution. 

The parameters imposing the biggest variations (see Eq.~\ref{massloss}) are the planetesimal eccentricity $e$, the width of the belt, and the factor $\gamma$ giving the composition
of planetesimals. Thus, a factor $\sim$ 10 variation can be explained if the parameter values are different from the fiducial values we picked. This can explain some 
discrepancies between our model predictions and the observations. A thorough study of each individual system would be needed to reduce these uncertainties but this is not the aim of this general study.

\begin{table}
    \caption{Parameters of the CO model that can be varied. We indicate the fiducial value picked for each parameter as well as a typical range of variations.}
\begin{center}

\begin{threeparttable}
\begin{tabular}{|l|c|c|}
  \toprule
  Parameters  & Fiducial & Range of \\
   & value & variation\\
  \midrule
L$_\star$ (L$_\odot$) & 10 & $10\%$\\
$R_0$ (au)& 85 & factor 2\\
L$_{\rm IR}$/L$_\star$ & $10^{-4}$ & $10\%$\\
${\rm d}r/r$ & 0.5 & 0.1-1.5\\
$e$ & 0.05 & 0.01-0.2\\
$\rho$ (kg/m$^3$) & 3000 & 1000-3500\\
$Q_D^\star$ (J/kg) & 500 & 100-1000\\
$t_{\rm ph}$ (yr) & 120 & factor 3\\
$\epsilon_{\rm CO}$ & 1 & factor 2\\
$\gamma$ (in $\%$) & 6 & $2-60$\\

  \bottomrule
\label{taberrorbar}
\end{tabular}
\end{threeparttable}

\end{center}

\end{table}

\subsection{CO succinct conclusion}
To conclude, using a simple model with $L_{\rm IR}$/$L_\star$, $R_0$, $L_\star$ and $d$ as free parameters, we are able to explain most CO observations to date. We also explain why CO was not easy to detect with single dish 
telescopes \citep[e.g.][]{2005MNRAS.359..663D,2011ApJ...740L...7M,2014AJ....148...47H,2015ApJ...814...42M,2016MNRAS.461.3910G}. Given a large sample of debris disc systems, we show
that ALMA will still detect CO over the next few years but one expects integration times longer than one hour to reach a large number of systems. In Fig.~\ref{fig1}, we show the part of the {$L_\star$, $R_0$} parameter space that should be avoided when looking for CO, and in Fig.~\ref{figCO} we study
the rest of the parameter space and give a way to calculate the predicted CO mass within each individual system and compare to detection thresholds. With our method, the whole parameter space is then studied and we show that the most important parameters (if one excludes the hatched zones plotted in
Fig.~\ref{fig1}) are the fractional luminosities $L_{\rm IR}$/$L_\star$ and the distance to Earth $d$ that have a quadratic dependence on CO mass. One should therefore look for CO with ALMA by picking systems with large IR-excesses, close to Earth and having $L_\star$ small enough and $R_0$ large enough not to lie
in the hatched exclusion areas in Fig.~\ref{fig1}. We also show that the parameters that are not observable directly that matter the most are the dynamical excitation of the disc, $\gamma$ and the belt width (see Fig.~\ref{figvarco}). 
The belt width is not known for unresolved debris discs. For systems that have their main belt resolved, the less extended, the better. The composition of planetesimals ($\gamma$) is also an important parameter as it provides the CO mass content in planetesimals. 

\section{Understanding carbon}\label{carbon}

In the same way as we calculated CO predictions from a simple analytical model, we will now do the same for carbon observations (CI and CII). To do so, we use the scenario presented in KWC16 where CO is input
within the system and photodissociates quickly into carbon and oxygen, which viscously spreads. As deduced from $\beta$ Pic observations, $\alpha$, which parameterises the viscous evolution, should be high and the corresponding
viscous timescale is $\sim 10^5$yr (see KWC16). We will assume that $\alpha=0.5$ throughout this paper, which sets the viscous timescale

\begin{equation}\label{viscous}
 t_\nu=R_0^2 \Omega/(\alpha c_s^2),
\end{equation}

\noindent where $\Omega$ is the orbital frequency and $c_s=\sqrt{R_g T/\mu}$, the sound speed fixed by the gas temperature $T$ (both estimated at $R_0$), with $R_g$ being the ideal gas
constant and $\mu$ the mean molecular mass of the carbon+oxygen fluid (assumed to be 14). Assuming steady state, one can then estimate
the amount of carbon within each system from the CO mass, as follows

\begin{equation}
\label{carbmass}
M_{\rm tot C}=0.43 \times \dot{M}_{\rm CO} t_\nu,
\end{equation}

\noindent where $0.43=12/28$ is the molar mass ratio between carbon and CO, and $\dot{M}_{\rm CO}$ is worked out using Eq.~\ref{COloss}. Thus, we can already conclude from Eq.~\ref{massloss} that a high carbon mass is favoured by a high fractional luminosity, a dynamically hot belt, a high $L_\star$ and a small belt width-to-distance ratio ${\rm d}r/r$. The blue hatched area
in Fig.~\ref{fig1} should still be avoided as all CO should be removed rapidly and no replenishment is possible over time (unless carbon or oxygen is produced through other less volatile molecules). On the contrary, systems that have a high carbon mass can be located in the green hatched zone in Fig.~\ref{fig1} as at steady state the carbon mass does not depend on $t_{\rm ph}$.

To compare our model to observations we must compute the carbon ionization fraction for each system to work out the CI and CII masses. To do so, we assume that the recombination rate $R_{\rm recomb}$ equals the ionization rate $R_{\rm ion}$. The total recombination rate 
(in m$^{-3}$s$^{-1}$) is dominated by CII recombination and is equal to

\begin{equation}
\label{recomb}
R_{\rm recomb}=\alpha_{R_C}(T) n_{\rm C_{II}} n_e,
\end{equation}

\noindent where $\alpha_{R_C}(T)$ is the recombination rate coefficient for CII that depends slightly on temperature and is taken from \citet{2006ApJS..167..334B}. $n_{\rm C_{II}}$ and $n_e$ are the number densities of CII and electrons respectively, which we assume are equal,
as in our model we assume that electrons are only produced when CI photoionizes into CII. The photoionization rate for carbon is

\begin{equation}
\label{rion}
R_{\rm ion}=n_{\rm C_{I}} \int_{\nu_{\rm ion}}^\infty \frac{4\pi J_\nu}{h \nu} \sigma_{\rm ion} \, {\rm d}\nu,
\end{equation}

\noindent where $n_{\rm C_{I}}$ is the neutral carbon number density, $\nu_{\rm ion}=11.26$eV is the smallest energy that can ionize CI and $\sigma_{\rm ion}$ is the carbon ionization cross section taken from \citet{2008CP....343..292V}. Also,
at steady state as the gas disc is an accretion disc, the surface density $\Sigma=\dot{M}_{CO}/(3 \pi \nu')$=2$\rho_g H$, with $\nu'$ the gas disc viscosity and $\rho_g$ the gas number density. To convert between $\Sigma$ and the particle number density $n$, we use $n=\rho_g/(\mu m_p)$ to find the carbon
number density at steady state

\begin{equation}
\label{nc}
n_{\rm C}=\frac{0.43 \dot{M}_{\rm CO} \Omega^2}{3 \pi \mu m_p c_s^3},
\end{equation}

\noindent where $m_p$ is the proton mass and we assumed $\alpha=0.5$. From this equation, one can compute the electron density anywhere in the system as $n_e=f n_{\rm C}$.
One can solve for the carbon ionization fraction $f$ 
by equating $R_{\rm ion}$ to $R_{\rm recomb}$ and using $f=n_{\rm C_{II}}/(n_{\rm C_{I}}+n_{\rm C_{II}})$ to find that

\begin{equation}
\label{ionfrac}
f=\frac{-R^*_{\rm ion}+\sqrt{{R^*_{\rm ion}}^2+4 R^*_{\rm ion} n_{\rm C} \alpha_{R_C}}}{2 n_{\rm C} \alpha_{R_C}},
\end{equation}

\noindent where $R^*_{\rm ion}=R_{\rm ion}/n_{\rm C_{I}}$, and $n_{\rm C}=n_{\rm C_{I}}+n_{\rm C_{II}}$. Hence the ionization fraction can be calculated knowing the radiation impinging on the disc $J_\nu$ and the carbon number density. There is also a slight temperature dependence through $\alpha_{R_C}$. 

We estimate the temperature $T_{\rm gas}$ at each location in the disc by equating cooling by the CII fine structure line and heating by CI photoionization and iterate a few times with the ionization fraction calculation until it converges. The calculations are described in appendix \ref{tempcalc} and we check that the
analytical formulation reproduces well previous published numerical simulations (see Fig.~\ref{figappen}) for the gas disc around $\beta$ Pictoris (see KWC16).

We are now able to compute the CI mass $M_{\rm tot C} (1-f)$ and the CII mass $M_{\rm tot C} f$ for any given system. $f$ depends on $R$ and is taken to be an average of the ionization fraction along $R$, by weighting with the surface density $R$-dependence. 
We thus find that $M_{\rm CII} \propto (L_{\rm IR} / L_\star)^{1/8} R_0^{1/2} \dot{M}_{\rm CO} T_{\rm gas}^{-1} f$, where $T_{\rm gas}$ is the gas temperature. 
As explained in the previous paragraph, we computed the temperature in the disc numerically but as a convenience for the reader, the following formula gives the CII mass (in M$_\oplus$) when assuming a power law for the gas temperature

\begin{equation}
\label{mc2}
M_{\rm CII}=\delta'  \left( \frac{L_\star}{L_\odot} \right)^{1/8} \left( \frac{R_0}{1{\rm au}} \right)^{1/2+\zeta} \left( \frac{\dot{M}_{\rm CO}}{0.23M_\oplus {\rm /Myr}} \right) f,
\end{equation}

\noindent where $\delta'=0.024/(\alpha T_0 (R_{T_0}/1 {\rm au})^\zeta)$. In this equation, the temperature profile is fixed to $T_{\rm gas}=T_0 (R/R_{T_0})^{-\zeta}$, where $\zeta$ is taken to be 0.5 in most studies, and we assume $\alpha=0.5$.
Substituting $\dot{M}_{\rm CO}$ from Eqs.~\ref{massloss} and \ref{COloss}, we find that $M_{\rm CII} \propto L_\star^{13/12} R_0^{1/6+\zeta} (L_{\rm IR}/L_\star)^{17/8} f$ for a fixed ${\rm d}r/r$.

Also, we can define the total CI mass with the same parameters using

\begin{equation}
\label{mc1}
M_{\rm CI}=M_{\rm CII} \frac{1-f}{f}.
\end{equation}

This set of equations will be used in the coming subsections to predict the CII and CI abundances in different systems. They can also be used theoretically to understand the system's parameters that matter the most to optimise the chances of finding new systems with gas.
  
\begin{figure*}
   \centering
   \includegraphics[width=16.5cm]{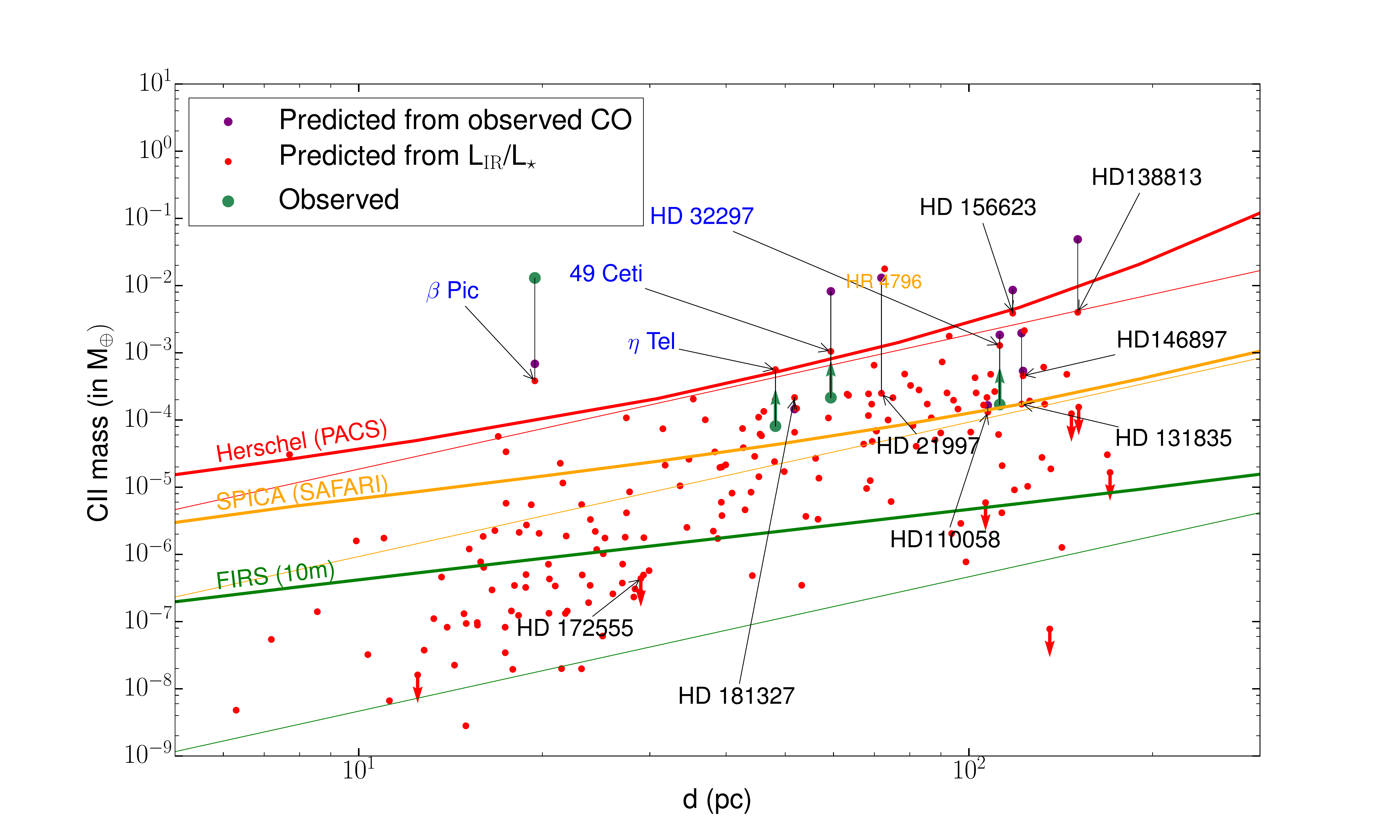}
   \caption{\label{figc2} CII mass (in M$_\oplus$) as a function of distance to Earth ($d$). Planetary systems with gas detections are labelled with their names. If CII is detected, the label is in blue (black otherwise). The CII mass lower limits worked out from observations are shown as green arrows,
and for $\beta$ Pic we show the mass derived from Herschel observations using \citet{2016MNRAS.461..845K} as a green point.
 The red points are predictions from our model. The red downward arrows show systems that are in the blue hatched area on Fig.~\ref{fig1}, which cannot keep CO trapped on solid bodies. The purple points show predictions from our model when the observed CO mass
is used rather than the CO mass predicted from $L_{\rm IR}/L_\star$. Detection limits at 5$\sigma$ in one hour are shown for Herschel/PACS (in red), 
SPICA/SAFARI (in orange) and FIRS (in green) for a 10m aperture. The thin lines are for LTE calculations and thick lines for more realistic NLTE calculations (using the same assumptions as described in section \ref{detect}).}
\end{figure*}


\subsection{Flux predictions for atoms}\label{fluxat}
When making predictions for CI, CII and OI masses, the conversion to fluxes is more complicated than with CO where the excitation is dominated by the CMB (see subsection \ref{fluxpr}). 
Since these lines are at shorter wavelengths, the dust radiation field can become dominant (see Fig.~\ref{figrad}). While the dust radiation field seen by the gas
is not easy to assess, especially as most of our targets are unresolved, it is well known for $\beta$ Pic. Thus, we use the same radiation field as found in Matr\`a et al., (in prep) for that system. For other targets, we fit each SED individually and use the ratio of the dust fluxes at 158, 610 and 63$\mu$m
to those of $\beta$ Pic to scale up or down the $\beta$ Pic radiation field and so get predictions for the fluxes of each individual system 
for CII, CI and OI (see Appendix \ref{NLTE}). Note that we also take account of the optical thickness of each line in our flux calculations using the method described in subsection \ref{thick}. 

\subsection{CII model predictions and results}

We start off by comparing our CII model predictions to Herschel observations. In Fig.~\ref{figc2}, we plot our predictions of CII mass as a function of distance to the star. To do so,
we use our analytical model using Eq.~\ref{carbmass} to get the total carbon mass and Eq.~\ref{ionfrac} to get the ionization fraction and then compute $M_{\rm CII}$. The predicted CII masses are shown as red points in Fig.~\ref{figc2}. We keep the same style as the CO plot (Fig.~\ref{figCO}), i.e systems with gas detected are labelled with their names. If in addition, they have
CII detected, the label is blue (not black). NLTE detection thresholds are computed using the same code as described to compute CO population levels in NLTE (see appendix \ref{NLTE} for details). We overplot the detection limits
at 5$\sigma$ in one hour in both LTE (thin lines) and NLTE (thick lines) for Herschel/PACS (in red) where we assumed the typical sensitivity reached for non detections \citep{2014A&A...565A..68R}, for SPICA/SAFARI (in orange) and for a future far-IR mission (such as FIRS) with a 10m aperture 
(in green) where the sensitivities are taken from the SPICA and FIR surveyor documentation\footnote{\url{https://firsurveyor.atlassian.net/}}. The assumed sensitivities are summarised in Table~\ref{tabsens}.
We compute the electron density as a function of the CII mass (using Eq.~\ref{nc}) with our model to compute the NLTE lines. 

We also compute the LTE limit in the same way as described in the previous section for CO.
We find that the transition between LTE and NLTE is at $M_{\rm CII} \sim 7 \times 10^{-5}$ M$_\oplus$.
Note that all the debris disc systems above the PACS detection threshold are most likely in LTE. The LTE Herschel detection threshold is, therefore, a good indicator of detectability unlike the case with CO. 

We also compute when the CII line becomes optically thick for an edge-on configuration. We use the same assumptions as for CO. Most of our 
predictions lie below the $\tau_\nu=1$ (edge-on) line located at $M_{\rm CII} \sim 2 \times 10^{-3}$M$_\oplus$. Therefore, we do not predict CII gas discs to be highly optically thick.
The NLTE lines are corrected for optical thickness (which is why the PACS sensitivity line steepens for large $d$). When the CII mass reaches $\sim 10^{-4}$M$_\oplus$, NLTE effects start affecting the Herschel and SPICA detection thresholds. 
New instruments as sensitive as FIRS will be able to detect gas discs in the NLTE regime.

Fig.~\ref{figc2} shows that our predictions for $\eta$ Tel, 49 Ceti, HD 32297 and $\beta$ Pic all lie above or close to the PACS detection threshold. The Herschel archive shows
that these targets were observed for at least 1.2 hours with PACS (HD 32997, which lies a bit below the threshold was observed for 2.6 hours). Note that the LTE mass detection threshold is for a temperature of 100K but
scales as $T^{1/2}$.

For CII detections, only lower limits on CII masses can be calculated from observations (as the excitation temperature is not known), which are represented as green arrows in Fig.~\ref{figc2}. Our predictions are all above these lower limits. 
For the CII mass in $\beta$ Pic, our prediction is about one order of magnitude below that observed. This can be explained
from KWC16, where it was found that the UV flux impinging on the disc should be higher than that assumed here to explain the CII observation. However, to be as general as possible in this study, we assume standard spectra for stars and a standard IRF.
This illustrates that our predictions have roughly order of magnitude uncertainty.

In addition to using our CO predictions to compute the CII masses, we show the results when using the observed CO masses $M_{\rm COobs}$ as purple points. We can then calculate $\dot{M}_{\rm CO}$ ($=M_{\rm COobs}/(t_{\rm ph} \epsilon_{\rm CO})$) in Eq.~\ref{carbmass} to 
make another prediction for CII. For most cases, a higher observed CO mass than predicted means a higher CII mass prediction. However, this is not straightforward for small CO mass variations (between the predicted and observed masses)
as increasing the CO mass will also decrease the ionization fraction (due to a higher carbon mass), which might be stronger than the increase in carbon mass. Also, increasing the CO mass can create more self-shielding, reducing the CO input rate and hence the CII mass. 
Using these new predictions does not change our previous conclusion that the 4 systems with CII
detected should have been detected. Three other systems, namely HD 21997, HD 138813 and HD 156623 cross the PACS detection threshold with these new predictions. 
However, as explained before, we cannot fit HD 21997 with a second generation scenario and this new prediction reinforces this idea
as CII was not detected by Herschel. Indeed, for a primordial gas origin, H$_2$ will shield CO photodissociation and carbon atoms will not be as abundant. As for  HD 138813 and HD 156623, the line was not observed with Herschel.

HR 4796 (labelled in orange) is the only star from the sample well above the PACS detection threshold in the optically thick region. However, we note that the flux will be lower than predicted because the CII line becomes optically thick.
We find that indeed, CII could not be detected with PACS but could be with SPICA. However, HR 4796 is an A0V star and the radiation pressure on CI is high
and could force CI to leave the system on dynamical timescales \citep{2006ApJ...643..509F}. We discuss radiation pressure effects in more detail in subsection \ref{radpre}.

Fomalhaut (the non-labelled red dot at 7.7pc) lies close to the detection threshold. Our flux prediction in Table~\ref{tabc2} is still lower than the published upper limit from PACS \citep[$2.2 \times 10^{-18}$ W/m$^2$,][]{2015A&A...574L...1C}. 
Other systems lie below the PACS detection threshold and, for those observed are indeed not detected.

New far-IR instruments such as SPICA or FIRS are needed to detect more CII gas discs. 
It is interesting to note that an instrument such as SPICA would increase our number of detections by a factor $\sim$ 7. For SPICA, targets with small CII masses will be
out of LTE. 
According to our flux predictions, SPICA could detect $\sim$ 25 new CII gas discs among which are Fomalhaut, HD 156623, HD 181327 and HR 4796.
Using the NLTE detection threshold for FIRS, we predict that it could detect CII in at least 100 systems. In Table~\ref{tabc2}, we provide a list of the most promising targets to look for CII with new missions.

\begin{table}
    \caption{List of the promising CII targets and their predicted masses, fluxes and observed fluxes.}
\begin{center}

\begin{threeparttable}
\begin{tabular}{|l|c|c|c|}
  \toprule
  Star's  & CII mass & $F_{\rm CII}$ \tiny{158$\mu$m} & $F_{\rm CII(obs)}$ \tiny{158$\mu$m} \\
  name & (M$_\oplus$) & (W/m$^2$) & (W/m$^2$) \\
  \midrule

Fomalhaut A & 3.1$\times 10^{-05}$ & 1.2$\times 10^{-18}$ & $<2.2\times 10^{-18a}$ \\
HD 86087 & 1.8$\times 10^{-03}$ & 1.2$\times 10^{-18}$ & - \\
HD 61005 & 2.0$\times 10^{-04}$ & 1.0$\times 10^{-18}$ & - \\
HD 156623 & 3.9$\times 10^{-03}$ & 8.3$\times 10^{-19}$ & -\\
HD 182681 & 6.5$\times 10^{-04}$ & 8.3$\times 10^{-19}$ & -\\
HR 4796 & 1.8$\times 10^{-02}$ & 8.2$\times 10^{-19}$ & $<2.2\times 10^{-18*}$\\
HD 131885 & 2.1$\times 10^{-03}$ & 7.6$\times 10^{-19}$ & -\\
HD 38678 & 1.2$\times 10^{-05}$ & 5.9$\times 10^{-19}$ & $<4.2\times 10^{-18*}$\\
HD 95086 & 7.3$\times 10^{-04}$ & 5.7$\times 10^{-19}$ & -\\
HD 138813 & 4.0$\times 10^{-03}$ & 5.7$\times 10^{-19}$ & -\\
HD 164249 & 2.0$\times 10^{-04}$ & 5.7$\times 10^{-19}$ & $<3.6\times 10^{-18b}$\\
HD 181327 & 2.2$\times 10^{-04}$ & 5.3$\times 10^{-19}$ & $<7.6\times 10^{-18b}$\\
HD 138965 & 4.8$\times 10^{-04}$ & 4.6$\times 10^{-19}$ & -\\
HD 221354 & 5.7$\times 10^{-05}$ & 4.6$\times 10^{-19}$ & -\\
HD 10647 & 3.4$\times 10^{-05}$ & 4.2$\times 10^{-19}$ & -\\
HD 124718 & 2.4$\times 10^{-04}$ & 3.8$\times 10^{-19}$ & -\\
HD 15745 & 2.3$\times 10^{-04}$ & 3.8$\times 10^{-19}$ & -\\
HD 191089 & 1.5$\times 10^{-04}$ & 3.8$\times 10^{-19}$ & -\\
HD 21997 & 2.5$\times 10^{-04}$ & 3.4$\times 10^{-19}$ & $<1.4\times 10^{-18*}$\\
HD 76582 & 1.3$\times 10^{-04}$ & 3.3$\times 10^{-19}$ & -\\
HD 192758 & 2.4$\times 10^{-04}$ & 3.0$\times 10^{-19}$ & -\\
HD 30447 & 3.3$\times 10^{-04}$ & 3.0$\times 10^{-19}$ & -\\
HD 6798 & 2.8$\times 10^{-04}$ & 2.7$\times 10^{-19}$ & -\\
HD 161868 & 7.4$\times 10^{-05}$ & 2.6$\times 10^{-19}$ & $<2.8\times 10^{-18*}$\\
HD 54341 & 4.3$\times 10^{-04}$ & 2.3$\times 10^{-19}$ & -\\
HD 111520 & 4.8$\times 10^{-04}$ & 2.0$\times 10^{-19}$ & -\\
HD 38206 & 2.1$\times 10^{-04}$ & 2.0$\times 10^{-19}$ & -\\
HD 106906 & 2.5$\times 10^{-04}$ & 2.0$\times 10^{-19}$ & -\\

  \bottomrule
\label{tabc2}
\end{tabular}
\begin{tablenotes}
		\footnotesize
   		\item $^a$ \citet{2015A&A...574L...1C}, $^b$ \citet{2014A&A...565A..68R}.
		\item $^*$ We obtained Herschel PACS CII data from the Herschel Science archive, and extracted spectra from Level 2 data products following the procedure described in the PACS Data Reduction Manual using 
HIPE v15.0.0. For pointed observations, spectra were obtained from the central 9.4'' spaxel (HD38678, HD164249, HR4796, HD161868) of the rebinned data cubes. For mapping observations (HD21997), we extracted a 
spectrum from the drizzle map by spatially integrating over spaxels over which continuum emission is detected. For all spectra, we first removed edge channels with extreme noise levels, then checked that the continuum level 
is in agreement with published measurements from the PACS photometer and subtracted it using a second order polynomial fit in spectral regions sufficiently distant from the CII line wavelength. As any emission present is expected 
to be spectrally unresolved at the resolution of the instrument (239 km/s), the 3$\sigma$ upper limits reported are simply the RMS of the final spectrum multiplied by the spectral resolution of the data.
  	\end{tablenotes}
\end{threeparttable}

\end{center}

\end{table}

\subsection{CI model predictions and results}

\begin{figure*}
   \centering
   \includegraphics[width=16.5cm]{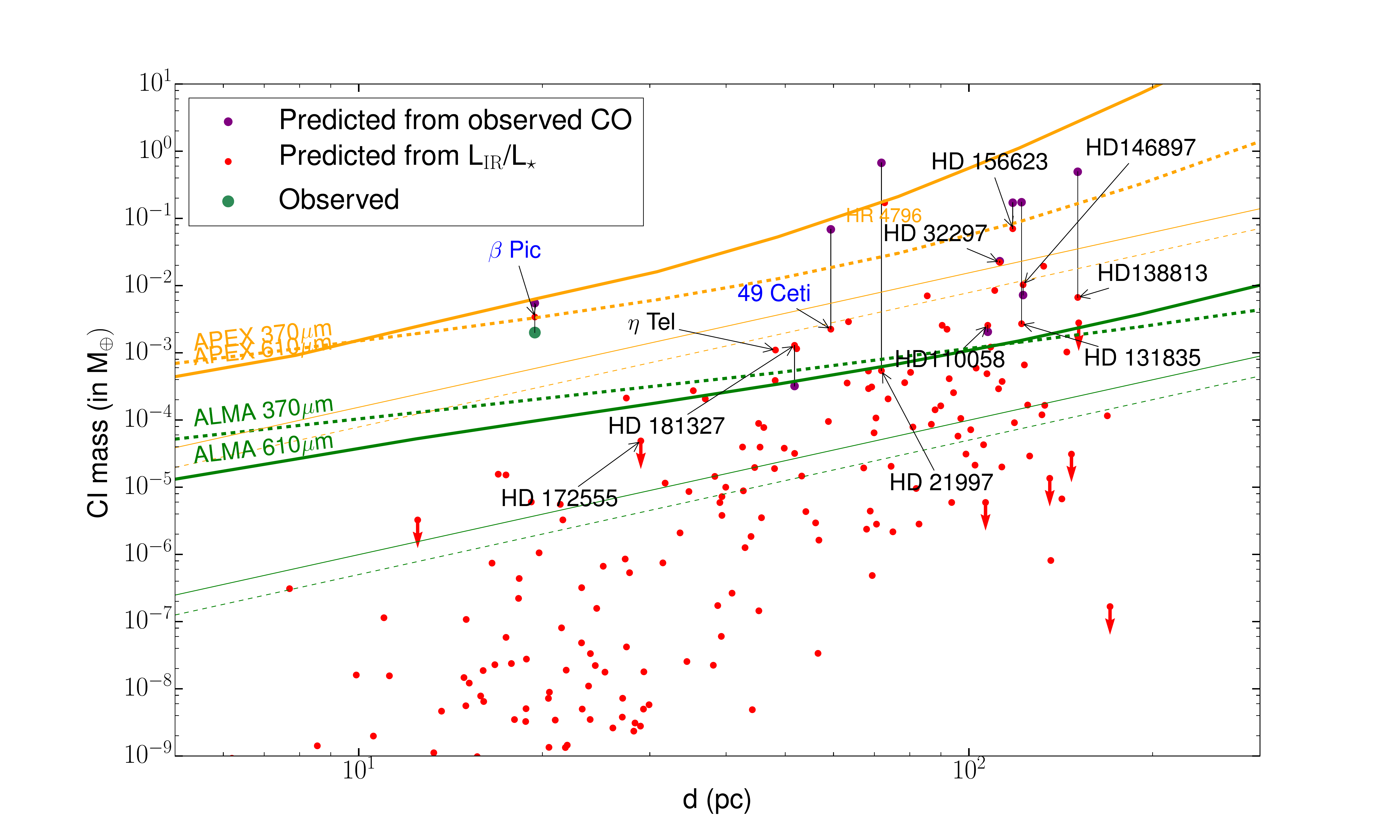}
   \caption{\label{figc1} CI mass (in M$_\oplus$) as a function of distance to Earth ($d$). Planetary systems with gas detections are labelled with their names. If CI is detected, the label is in blue (black otherwise). The CI mass for $\beta$ Pic,
 derived from \citet{2016MNRAS.461..845K} is shown as a green point.
 The red points are predictions from our model. The red downward arrows show systems that are in the blue hatched area on Fig.~\ref{fig1}, which cannot keep CO trapped on solid bodies. The purple points show predictions from our model when the observed CO mass
is used rather than the CO mass predicted from $L_{\rm IR}/L_\star$. Detection limits at 5$\sigma$ in one hour are shown for APEX (in orange) and ALMA (in green) at 370 (dotted) and 610 microns (solid). The thin lines are for LTE calculations and thick lines
for more realistic NLTE calculations (using the same assumptions as described in section \ref{detect}).}
\end{figure*}


CI has only been observed around $\beta$ Pic and 49 Ceti in absorption with the HST/STIS \citep{2000ApJ...538..904R,2014ApJ...796L..11R}. APEX has only provided upper limits so far (KWC16). 
We here investigate whether ALMA is likely to detect CI around other debris disc hosts, and which are the favoured systems in which to search for it.

In Fig.~\ref{figc1}, we repeat the same procedure as for CO and CII. We also compute the critical density for the CI to be in LTE and find that $M_{\rm CI} \sim 3 \times 10^{-2}$M$_\oplus$ is required. 
This is high enough that most detections that can be made with APEX and ALMA will be in NLTE. We computed the population levels in NLTE in Eq.~\ref{tau} to work out the $\tau_\nu=1$ line. For an edge-on configuration, we find that $M_{\rm CI} \sim 3 \times 10^{-4}$M$_\oplus$. 
Thus, the CI line is likely to be optically thick for the most distant systems with CO detected. For this reason, the CI detection limits increase faster than distance squared beyond some distance, meaning that these systems should not be detected with APEX even for the most CO-rich debris discs. 
Only $\beta$ Pic, 49 Ceti and HD 21997 for the predictions from the observed CO in purple, and HR 4796 are close to the APEX detection threshold (though note that HD 21997 is thought to be made of primordial gas). 
For other systems, only ALMA will be able to offer detections as can be seen from our flux predictions in Table~\ref{tab3}. 

One can see that the CI mass predicted for $\beta$ Pic in KWC16, shown as a green point in Fig.~\ref{figc1}, is well below the APEX detection threshold. 
Indeed, CI in $\beta$ Pic was not detected with APEX (20 minutes on source, KWC16). Similarly to CII, we expect the flux impinging on the $\beta$ Pic gas disc to be higher
than assumed here (as was found in KWC16\footnote{Here, we emphasise that this is a specific feature of $\beta$ Pic, probably due to its high stellar activity as explained in KWC16. However, we expect the flux impinging on other debris discs to be closer to the sum of the standard interstellar and stellar radiation fields.
}), therefore, the CI mass will go down and our $\beta$ Pic prediction will in reality be closer to the observation and even farther below the 20 minute APEX threshold (not shown here but $\sim$ 1.7 times higher than the one hour line). 
Our flux prediction of $2.3 \times 10^{-19}$ W/m$^2$ (even without increasing the impinging radiation) is close to the APEX upper limit from KWC16 and ALMA should detect CI easily in this system.


From our sample of debris disc stars, we find that with ALMA we could detect at least 30 systems (at 5$\sigma$ in an hour) that are listed in Table~\ref{tab3}. By pushing the integration time to 5 hours,
we could reach $\sim$ 45 systems. We also plot the detection limits at 370$\mu$m, which correspond to the higher CI transition. This transition can be more sensitive with APEX for temperatures higher than $\sim 65$K. 

Our flux predictions in Table~\ref{tab3} show that $\beta$ Pic and 49 Ceti, the two systems with CI detected, are indeed among the three most favourable targets. Systems such as $\eta$ Tel, HD 156623, HD 172555, HD 32297, HD 181327, HD 110058 that have detected gas should
be searched for CI first, as a combination of CO+CI or CII+CI (for $\eta$ Tel) can provide much more information on the systems (e.g. value of the viscosity $\alpha$, ionization fraction).

Therefore, we predict that ALMA observations of CI are a promising way to detect secondary gas in debris discs.
Also, thanks to ALMA's very high-resolution, it will be possible to explore the inner parts of planetary systems and might provide a new complementary picture compared to dust observations. 
These CI observations could be used to study the gas distribution in the inner regions of planetary systems, which might trace the location of new inner planets (if structures are observed in these atomic gas discs).
Also, the discovery of more of these new atomic gas discs will enrich our knowledge of the gas dynamics and more values for $\alpha$ (which parameterises the viscosity) could be calculated and compared to the MRI theory \citep[e.g.][]{2016MNRAS.461.1614K}. 

\begin{table}
    \caption{List of ALMA promising CI targets and their predicted masses and fluxes.}
\begin{center}

\begin{threeparttable}
\begin{tabular}{|l|c|c|}
  \toprule
  Star's  & CI mass & $F_{\rm CI}$ \tiny{610$\mu$m} \\
  name & (M$_\oplus$) & (W/m$^2$) \\
  \midrule

$\beta$ Pic & 3.4$\times 10^{-03}$ & 2.3$\times 10^{-19}$ \\
HR 4796 & 1.7$\times 10^{-01}$ & 4.9$\times 10^{-20}$ \\
49 Ceti & 2.2$\times 10^{-03}$ & 4.9$\times 10^{-20}$ \\
HD 156623 & 7.0$\times 10^{-02}$ & 4.3$\times 10^{-20}$ \\
$\eta$ Tel & 1.1$\times 10^{-03}$ & 2.1$\times 10^{-20}$ \\
HD 138813 & 6.7$\times 10^{-03}$ & 2.0$\times 10^{-20}$ \\
HD 21997 & 5.4$\times 10^{-04}$ & 1.5$\times 10^{-20}$ \\
HD 131835 & 2.7$\times 10^{-03}$ & 1.4$\times 10^{-20}$ \\
HD 191089 & 1.2$\times 10^{-03}$ & 1.4$\times 10^{-20}$ \\
HD 15745 & 2.9$\times 10^{-03}$ & 1.3$\times 10^{-20}$ \\
HD 172555 & 4.9$\times 10^{-05}$ & 1.2$\times 10^{-20}$ \\
HD 32297 & 2.2$\times 10^{-02}$ & 9.6$\times 10^{-21}$ \\
HD 181327 & 1.3$\times 10^{-03}$ & 9.6$\times 10^{-21}$ \\
HD 114082 & 7.0$\times 10^{-03}$ & 7.9$\times 10^{-21}$ \\
HD 95086 & 2.6$\times 10^{-03}$ & 7.9$\times 10^{-21}$ \\
HD 86087 & 4.1$\times 10^{-04}$ & 7.6$\times 10^{-21}$ \\
HD 61005 & 2.7$\times 10^{-04}$ & 6.8$\times 10^{-21}$ \\
HD 106906 & 2.2$\times 10^{-03}$ & 6.5$\times 10^{-21}$ \\
HD 129590 & 1.9$\times 10^{-02}$ & 6.5$\times 10^{-21}$ \\
HD 107146 & 2.1$\times 10^{-04}$ & 6.4$\times 10^{-21}$ \\
HD 117214 & 8.5$\times 10^{-03}$ & 6.3$\times 10^{-21}$ \\
HD 146897 & 1.0$\times 10^{-02}$ & 5.7$\times 10^{-21}$ \\
HD 164249 & 3.9$\times 10^{-04}$ & 5.6$\times 10^{-21}$ \\
HD 110058 & 2.6$\times 10^{-03}$ & 5.4$\times 10^{-21}$ \\
HD 131885 & 6.6$\times 10^{-04}$ & 4.9$\times 10^{-21}$ \\
HD 221853 & 5.4$\times 10^{-04}$ & 4.4$\times 10^{-21}$ \\
HD 121191 & 2.8$\times 10^{-03}$ & 3.8$\times 10^{-21}$ \\
HD 69830 & 3.3$\times 10^{-06}$ & 3.2$\times 10^{-21}$ \\
HD 170773 & 2.1$\times 10^{-04}$ & 2.7$\times 10^{-21}$ \\
HD 124718 & 3.6$\times 10^{-04}$ & 2.6$\times 10^{-21}$ \\
HD 38678 & 3.3$\times 10^{-06}$ & 2.0$\times 10^{-21}$ \\
HD 182681 & 6.4$\times 10^{-05}$ & 1.8$\times 10^{-21}$ \\
HD 35841 & 5.9$\times 10^{-04}$ & 1.8$\times 10^{-21}$ \\
HD 106036 & 7.2$\times 10^{-05}$ & 1.8$\times 10^{-21}$ \\

  \bottomrule
\label{tab3}
\end{tabular}
\end{threeparttable}

\end{center}

\end{table}

\subsection{Atomic mass variation when changing parameters}\label{atommas}

\begin{figure*}
   \centering
   \includegraphics[width=18.5cm]{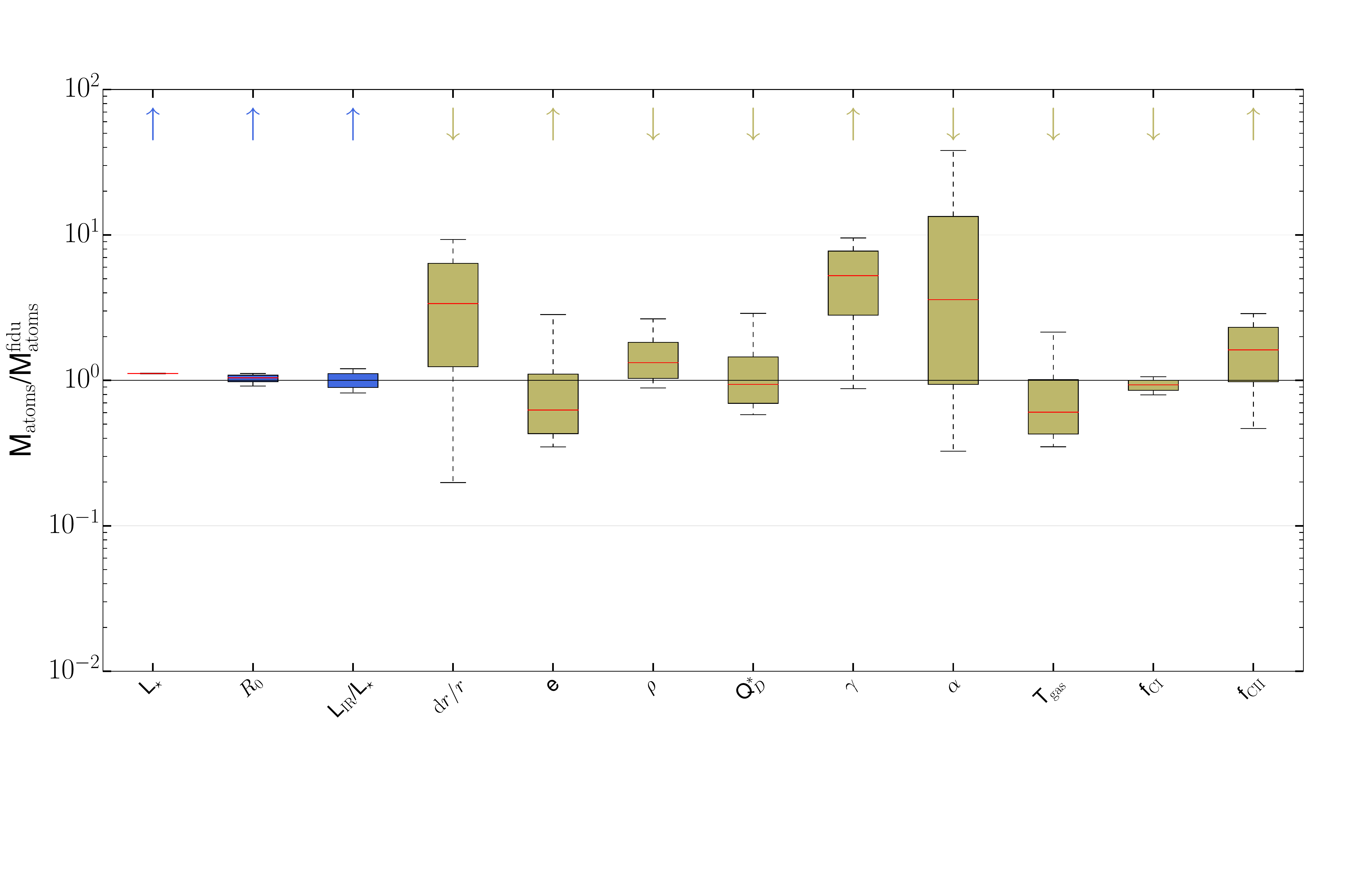}
   \caption{\label{figvarc} CI, CII or OI mass variations whilst varying parameters one by one. The fiducial values and amplitude of variations of the different parameters are given in Table~\ref{taberrorbar2}. The blue boxes are for parameters where we have a good observational handle and only error bars on the predicted 
values are taken into account. For fawn boxes, there are more uncertainties and we allow for a large physical variation for each parameter. Box sizes show where 50\% of the distribution is located, while the whiskers contain 95\% of the distribution. The red line is the median value. We indicate with an upwards or
downwards arrow the sense of variation for $M_{\rm atoms}$ if a given parameter is increased.}
\end{figure*}

\begin{table}
    \caption{Parameters of the CI, CII, OI model that can be varied. We indicate the fiducial value picked for each parameter as well as a typical range of variations.}
\begin{center}

\begin{threeparttable}
\begin{tabular}{|l|c|c|}
  \toprule
  Parameters  & Fiducial & Range of \\
   & value & variation\\
  \midrule
L$_\star$ (L$_\odot$) & 10 & $10\%$\\
$R_0$ (au)& 85 & factor 2\\
L$_{\rm IR}$/L$_\star$ & $10^{-4}$ & $10\%$\\
${\rm d}r/r$ & 0.5 & 0.1-1.5\\
e & 0.05 & 0.01-0.2\\
$\rho$ (kg/m$^3$) & 3000 & 1000-3500\\
$Q_D^\star$ (J/kg) & 500 &100-1000\\
$\gamma$ (in $\%$) & 6 & $2-60$\\
$\alpha$ (log) & 0.5 & 0.01-2\\
T$_{\rm gas}$ (K) & 10 & factor 3\\
f  & 0.1 & factor 3\\
  \bottomrule
\label{taberrorbar2}
\end{tabular}
\end{threeparttable}

\end{center}

\end{table}

We here study the impact on our atomic mass predictions when varying parameters. Fig.~\ref{figvarc} shows the variations expected when parameters vary from their fiducial values within the allowed range (see Table~\ref{taberrorbar2}). The upwards and downwards arrows show the direction
of a change in atomic mass when a given parameter is increased.
Compared to CO, some new parameters come into play.
Indeed, the atomic masses depend on $\alpha$, T$_{\rm gas}$ and $f$ but do not depend on $t_{\rm ph}$ or $\epsilon_{\rm CO}$. 

One can see that for atoms, the most important parameters are $\alpha$, the belt width and $\gamma$. By varying these parameters, one can account for a factor $\sim$ 10
in either direction between our predictions and observations. These variations are the same for CI, CII or OI except for the last two parameters $f_{\rm CI}$ and $f_{\rm CII}$ listed in Fig.~\ref{figvarc}, which are the variations implied by a change of $f$ on CI and CII masses, respectively.

\section{Understanding oxygen}\label{ox}

\begin{figure*}
   \centering
   \includegraphics[width=16.5cm]{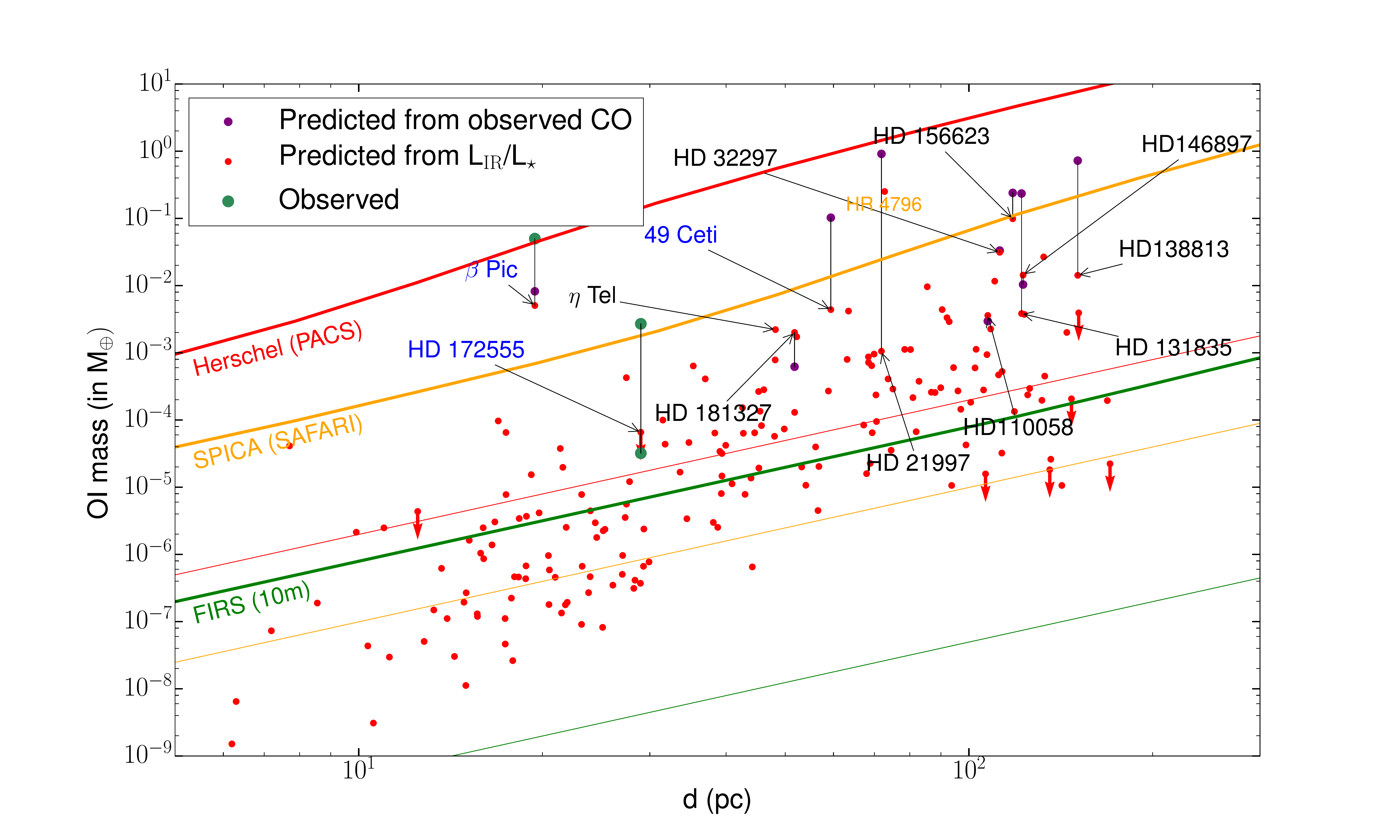}
   \caption{\label{figo1} OI mass (in M$_\oplus$) as a function of distance to Earth ($d$). Planetary systems with gas detections are labelled with their names. If OI is detected, the label is in blue (black otherwise). The OI mass for $\beta$ Pic, 
 derived from PACS observations \citep{2016MNRAS.461..845K,2016A&A...591A..27B} is shown as a green point as well as the observed mass for HD 172555 considering the flux observed with PACS \citep{2012A&A...546L...8R}. The second green point (at a lower mass) for HD 172555
considers that some extra hydrogen comes off the grains at the same time as CO is released (see section \ref{disc}).
The red points are predictions from our model. The red downward arrows show systems that are in the blue hatched area on Fig.~\ref{fig1}, which cannot keep CO trapped on solid bodies. The purple points show predictions from our model when the observed CO mass
is used rather than the CO mass predicted from $L_{\rm IR}/L_\star$. Detection limits at 5$\sigma$ in one hour are shown for Herschel/PACS (in red), SPICA/SAFARI (in orange) and FIRS (in green) for a 10m aperture. The thin lines are for LTE calculations and thick lines
for more realistic NLTE calculations (using the same assumptions as described in section \ref{detect}).}
\end{figure*}

We proceed in the same way as described in previous sections to produce Fig.~\ref{figo1}. 
We assume that oxygen stays neutral as its ionization potential is 13.6eV and UV photons with such high energies are depleted around A type or later-type stars \citep{2010ApJ...720..923Z}. OI
was detected in absorption with HST around 49 Ceti. Herschel only detected OI around $\beta$ Pic
and HD 172555. We know that HD 172555 is in the blue hatched area in Fig.~\ref{fig1} and so grain temperatures might be too high to maintain CO on solids. In this particular system, it could be that OI is created
from SiO photodissociation or evaporation of O-rich refractories rather than CO \citep{2009ApJ...701.2019L}. However, we decided to show our model prediction
for OI if CO could survive on solid bodies in HD 172555. We notice that HD 172555 stands out compared to other systems with gas detected as the OI mass predicted is the lowest. 

The LTE limit for OI (63 $\mu$m)
is at 38 M$_\oplus$. Indeed, the OI critical electron density is very high because the Einstein A coefficients (and hence spontaneous decays) are higher than other cases. Therefore, the whole parameter space shown in Fig.~\ref{figo1} is out of LTE. LTE is a very bad
approximation for OI and should not be used. The LTE/NLTE detection thresholds are shown in red for PACS, orange for SAFARI and green for FIRS (10m). The thick NLTE lines are $\sim$3 orders of magnitude less sensitive than LTE. 
Therefore, one needs a 1000 times higher OI mass (compared to LTE) to detect a system that is out of LTE. 

The PACS NLTE detection threshold shows that indeed the detection of OI around $\beta$ Pic is above the one-hour detection limit. We note however that our prediction for $\beta$ Pic (in red) lies below the detection threshold. Indeed, in KWC16, we found by fitting the OI PACS spectrum that the OI mass 
needed some extra oxygen coming from water in addition to the oxygen coming from CO to fully  explain the observed flux ($1.7 \times 10^{-17}$ W/m$^2$) with a total oxygen mass of $5 \times 10^{-2}$ M$_\oplus$ (green point
in Fig.~\ref{figo1}). Including oxygen coming from water photodissociation in other targets would increase their OI mass and the extra HI may act as an extra collider together with electrons to make the OI line easier to detect. Our flux prediction
for $\beta$ Pic would change from $1.1 \times 10^{-18}$ W/m$^2$ to $1.4 \times 10^{-17}$ W/m$^2$ when adding extra water (see Table \ref{tabo1b}). We discuss this idea further in section \ref{disc}.
We find that HR 4796 lies well below the PACS detection limit. OI observations for this system were attempted with PACS but led to no detection, as predicted by our model. This system is however the most promising as shown in Table~\ref{tabo1b} which shows our flux predictions
for systems that could be observed with SPICA. Other systems lie below the PACS detection threshold. This is once again consistent with observations that have been made so far.

For HD 172555, we recomputed a new mass from the observed PACS flux equal to $9 \times 10^{-18}$ W/m$^2$ \citep{2012A&A...546L...8R}, taking into account NLTE effects and optical thickness of the line. We find an OI mass of $3 \times 10^{-3}$ M$_\oplus$ (green point on Fig.~\ref{figo1}). We also
compute the observed mass if some extra water (and then hydrogen) comes off the grain while releasing CO (see subsection \ref{disc}) and find an OI mass of $3 \times 10^{-5}$ M$_\oplus$ (second green point for HD 172555), which is closer to our prediction.
From Table~\ref{tabo1b}, we see that we also predict that the OI flux for HD 172555 is $\sim 3 \times 10^{-19}$ W/m$^2$, which is below the PACS sensitivity and $\sim$ 2.5$\times 10^{-17}$ W/m$^2$ with extra water, which is detectable with PACS (see the discussion). 
The flux prediction for this system is high given its low predicted mass in Fig.~\ref{figo1}.
Because HD 172555 parent belt is within a few au,
the electron density will be much higher than assumed when plotting the detection threshold in Fig.~\ref{figo1}, making this oxygen line much closer to the LTE regime and the new detection threshold much closer to the thin lines shown in Fig.~\ref{figo1}.
While HD 172555 is not predicted to have detectable levels of OI, it is the third highest flux prediction and it could be that (as in $\beta$ Pic), some water
is also released together with CO, which would boost our prediction and explain the PACS detection (see section \ref{disc}). OI could also be produced from gas released from refractory elements as suggested in \citet{2009ApJ...701.2019L}.

Fig.~\ref{figo1} also shows that the new far-IR instrument SAFARI on SPICA may lead to a few more detections. The orange line is 20 times more sensitive than PACS and would obtain $\sim$ 3 new detections if we integrate on source 5 hours. More
detections would be possible if water is released together with CO.
 A mission such as FIRS could detect debris disc stars with OI. The NLTE FIRS detection threshold (thick green) is $\sim$ 200 times more sensitive than the NLTE SAFARI (thick orange) line and could enable detection of $\sim$ 35 new systems.

Detections of OI with SPICA would be a great way to assess the amount of water in these systems and see how much it contributes to the overall OI flux. In Table~\ref{tabo1b}, we provide a list of the most promising targets that should be looked for
with any new facility that can target the OI 63$\mu$m line.

\begin{table*}
    \caption{List of SPICA promising targets to look for OI and their predicted masses and fluxes without (2 first columns) and with extra water released together with CO (2 following columns). The last column gives observed fluxes or upper limits when systems were observed with Herschel.}
\begin{center}

\begin{threeparttable}
\begin{tabular}{|l|c|c|c|c|c|}
  \toprule
  Star's  & OI mass & $F_{\rm OI}$ \tiny{63$\mu$m} & OI mass (with H$_2$O) & $F_{\rm OI}$ {\tiny 63$\mu$m} (with H$_2$O) & $F_{\rm OI(obs)}$ {\tiny 63$\mu$m}\\
  name & (M$_\oplus$) & (W/m$^2$) & (M$_\oplus$) & (W/m$^2$) & (W/m$^2$) \\
  \midrule

$\beta$ Pic & 5.1$\times 10^{-03}$ & 1.1$\times 10^{-18}$ & 1.5$\times 10^{-02}$ & 1.4$\times 10^{-17}$ & 1.7$\times 10^{-17a}$ \\
HR 4796 & 2.5$\times 10^{-01}$ & 3.7$\times 10^{-19}$ & 7.5$\times 10^{-01}$ & 4.4$\times 10^{-18}$ & $<4.7\times 10^{-18b}$ \\
HD 172555 & 6.6$\times 10^{-05}$ & 3.1$\times 10^{-19}$ & 2.0$\times 10^{-04}$ & 2.5$\times 10^{-17}$ & $9.2 \pm 2.4 \times 10^{-18c}$ \\
HD 121191 & 3.9$\times 10^{-03}$ & 9.9$\times 10^{-20}$ &1.2$\times 10^{-02}$ & 1.7$\times 10^{-18}$ & - \\
$\eta$ Tel & 2.2$\times 10^{-03}$ & 8.9$\times 10^{-20}$ &6.6$\times 10^{-03}$ & 1.6$\times 10^{-18}$ & 6.2$\times 10^{-18d}$ \\
Fomalhaut A & 4.1$\times 10^{-05}$ & 5.9$\times 10^{-20}$& 1.2$\times 10^{-04}$ & 2.4$\times 10^{-19}$ & 1.0$\times 10^{-17e}$ \\
HD 138923 & 1.6$\times 10^{-05}$ & 4.3$\times 10^{-20}$ &4.7$\times 10^{-05}$ & 2.7$\times 10^{-18}$ & - \\
HD 156623 & 9.9$\times 10^{-02}$ & 4.2$\times 10^{-20}$ &3.0$\times 10^{-01}$ & 3.0$\times 10^{-18}$ & - \\
49 Ceti & 4.4$\times 10^{-03}$ & 3.5$\times 10^{-20}$ &1.3$\times 10^{-02}$ & 3.2$\times 10^{-18}$ & $< 1.1 \times 10^{-17f}$ \\
HD 106036 & 1.8$\times 10^{-04}$ & 1.3$\times 10^{-20}$ &5.5$\times 10^{-04}$ & 1.1$\times 10^{-18}$ & - \\
HD 138813 & 1.4$\times 10^{-02}$ & 1.2$\times 10^{-20}$ &4.3$\times 10^{-02}$ & 1.2$\times 10^{-18}$ & $< 1.3 \times 10^{-17g}$ \\
HD 181327 & 2.0$\times 10^{-03}$ & 1.1$\times 10^{-20}$ &6.0$\times 10^{-03}$ & 5.5$\times 10^{-19}$ & $< 8.2 \times 10^{-18d}$ \\


  \bottomrule
\label{tabo1b}
\end{tabular}
\begin{tablenotes}
		\footnotesize
   		\item $^a$ KWC16, $^b$ \citet{2013A&A...555A..67R}, $^c$ \citet{2012A&A...546L...8R}, $^d$ \citet{2014A&A...565A..68R}, $^e$ \citet{2015A&A...574L...1C}, $^f$ \citet{2013ApJ...771...69R}, $^g$ \citet{2013A&A...558A..66M}.
  	\end{tablenotes}
\end{threeparttable}

\end{center}

\end{table*}

\section{Discussion}\label{disc}

\subsection{Radiation pressure on CI, CII and OI}\label{radpre}

Here, we discuss the effect of having an accretion disc which extends all the way to the star on the radiation pressure force felt by atoms.

\begin{figure}
   \centering
   \includegraphics[width=8.5cm]{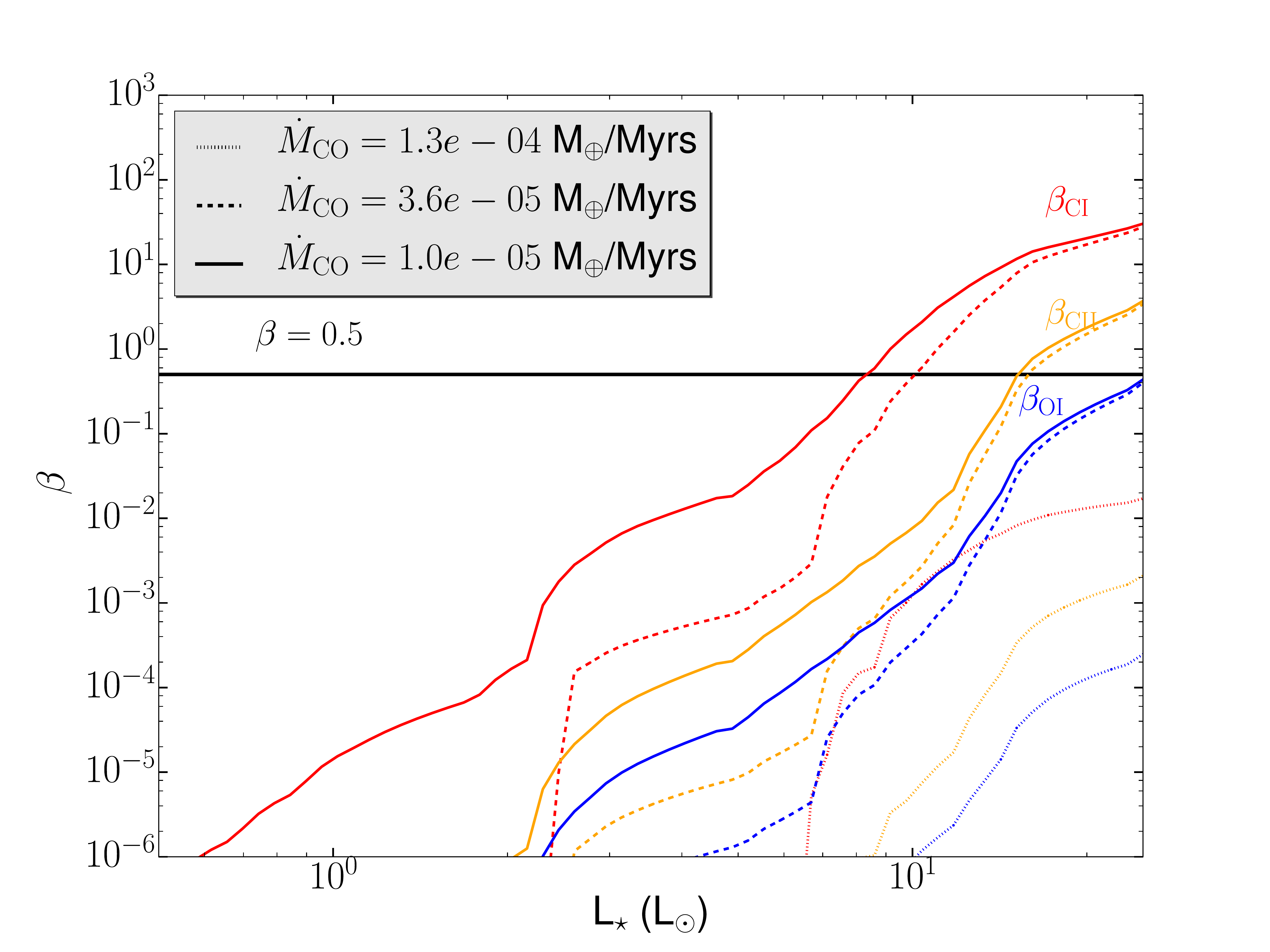}
   \caption{\label{figbeta} $\beta$ as a function of L$_\star$ computed for three different atomic species CI (in red), CII (in yellow) and OI (in blue) for different CO input rates $10^{-5}$ M$_\oplus$/Myrs (solid line), $3.6 \times 10^{-5}$ (dashed) and $1.3 \times 10^{-4}$ (dotted). 
The black line shows the location of $\beta=0.5$, above which atoms are unbound.}
\end{figure}

In Fig.~\ref{figbeta}, we show how $\beta$, the radiation pressure force relative to gravity, varies with the CO input rate $\dot{M}_{\rm CO}$ and $L_\star$ for different species. The radiation pressure on atoms comes from the star as the IRF is assumed to be isotropic so has a zero net overall effect on
radiation pressure. For high enough $\dot{M}_{\rm CO}$, the accretion disc will become optically thick to UV radiation in the radial direction, decreasing the effectiveness of the star's radiation pressure. 

For $\beta$ Pic, it is predicted that with sufficiently high CII mass in the system, metals will brake due to Coulomb collisions with CII, which is not affected by radiation pressure \citep{2006ApJ...643..509F}. However, for early stellar types (earlier than A5V), $\beta_{\rm C_{I}}$, the effective $\beta$ for CI, can become greater than 0.5. Therefore,
without any shielding from the star, CI would be blown out from the system. Carbon could not be kept in its ionized form CII as it constantly transforms into CI on an ionization timescale but rather all the carbon would be blown out. In Fig.~\ref{figbeta}, we quantify the luminosity at which this transition happens and also how much mass is required to stop CI from being blown out.
To do so, we compute $\beta$ as \citep{2006ApJ...643..509F}

\begin{equation}
\label{betac1}
\beta=\frac{R^2}{8 \pi c^2 G M_\star m} \sum_{i<j} \frac{g_j}{g_i} A_{ji} \lambda^4_{ij} F_\lambda,
\end{equation}

\noindent where $M_\star$ is the stellar mass and $m$ is the mass of the considered atom (here, carbon or oxygen) for which $\beta$ is computed. $g_j$ and $g_i$ are the j-th and i-th statistical weights, $A_{ji}$ the Einstein A coefficient corresponding to the j to i transition, and $\lambda_{ij}$
the transition wavelength (all the transitions were downloaded  from the NIST database\footnote{\url{https://www.nist.gov/pml/atomic-spectra-database}}). Note that $\beta$ does not depend upon R as the stellar flux $F_\lambda$ and gravity both scale as $R^{-2}$.

We find that without shielding, CI should start to be blown out in systems with stellar luminosity greater than $\sim$ 8 L$_\odot$. However, for a relatively small value of $\dot{M}_{\rm CO}$, $\beta_{\rm CI}$
goes below 0.5 even around these highly luminous stars because of self-shielding.  We predict that in all systems with $\dot{M}_{\rm CO} > 10^{-4}$ M$_\oplus$/Myrs, CI can be protected from being blown out. We also predict that without shielding CII would be blown out for systems with L$_\star >$15 L$_\odot$ but OI would stay bound up to 25 L$_\odot$. 

According to our predictions, a system with a CO mass input rate $\sim$ 1000 times smaller than in $\beta$ Pic is enough to keep CI or CII from being blown out. Thus, it is likely that this effect will only affect systems with very low CO mass input rates. We can check on Fig.~\ref{figCO}
that a system with one thousandth of the $\beta$ Pic mass would not be detectable with ALMA.


\subsection{Caveats}\label{cavea}

{\it Analytical version of the code:} The semi-analytical model presented here has a few caveats. First of all, we assume Eq.~\ref{massloss} to compute the mass lost through the cascade. This equation is only valid at steady state for a typical -3.5 size distribution. Some more refined numerical models
could be used to derive the lost mass using more realistic particle size distributions \citep[e.g.][]{2011CeMDA.111....1W} or departing from the steady state assumption \citep[e.g.][]{2007A&A...472..169T,2008ApJ...673.1123L,2015A&A...573A..39K}. In this equation, some parameters are not directly accessible to observers such as the 
planetesimal eccentricity, their bulk density or their collisional strength Q$_D^*$. Also, in Eq.~\ref{COloss}, we assume that a fraction $\gamma$ of the dust mass is converted into CO. For instance, varying $\gamma$ can give us a way to fit the prediction with the observation and thus constrain the
amount of CO on planetesimals. However, through Figs.~\ref{figvarco}
and \ref{figvarc}, we were able to quantify the impact of each parameter variation. We concluded that for a given system, the predicted mass can vary by a factor 10. This can explain some of the differences between observations and predictions and ultimately could lead to constraints on some of these
free parameters.

{\it Cooling/heating:} Another assumption is that the only coolant is the CII fine structure line and that the only heating mechanism is CI photoionization. For the low temperatures expected
in debris discs and given the amount of dust in these systems, this is likely to be a good approximation. Indeed, as shown in KWC16, one needs very dusty discs for photoelectric heating to dominate over CI photoionization. To be more specific, one needs dustier systems than $\beta$
Pic with an optical depth $\gtrsim 10^{-2}$, which is never the case for debris discs (by definition). Thus, this assumption is likely to always be valid.  
Cooling by the OI fine structure line at 63.2$\mu$m may also be important. However, when carbon is present, and for the range of temperatures expected in these gas discs, the CII line is always the dominant coolant as shown in \citet{2010ApJ...720..923Z} and using Cloudy simulations in KWC16.
It is mainly due to the OI line being in strong NLTE in debris discs.

{\it Extra water released together with CO:} In this paper, we assumed that there was no extra water released at the same time as CO. However, if CO (+CO$_2$ which contributes to providing more CO when photodissociating) is trapped in water ice, a certain amount of H$_2$O may also be released in the process (see KWC16). 
Water photodissociates a lot faster than CO and this will add more oxygen in the system, but also excite the OI line more due to extra collisions with neutral hydrogen. The amount of water released at the same time as CO is not known yet and thus
the amount of extra colliders or extra oxygen can only be assumed.

\begin{figure*}
   \centering
   \includegraphics[width=16.5cm]{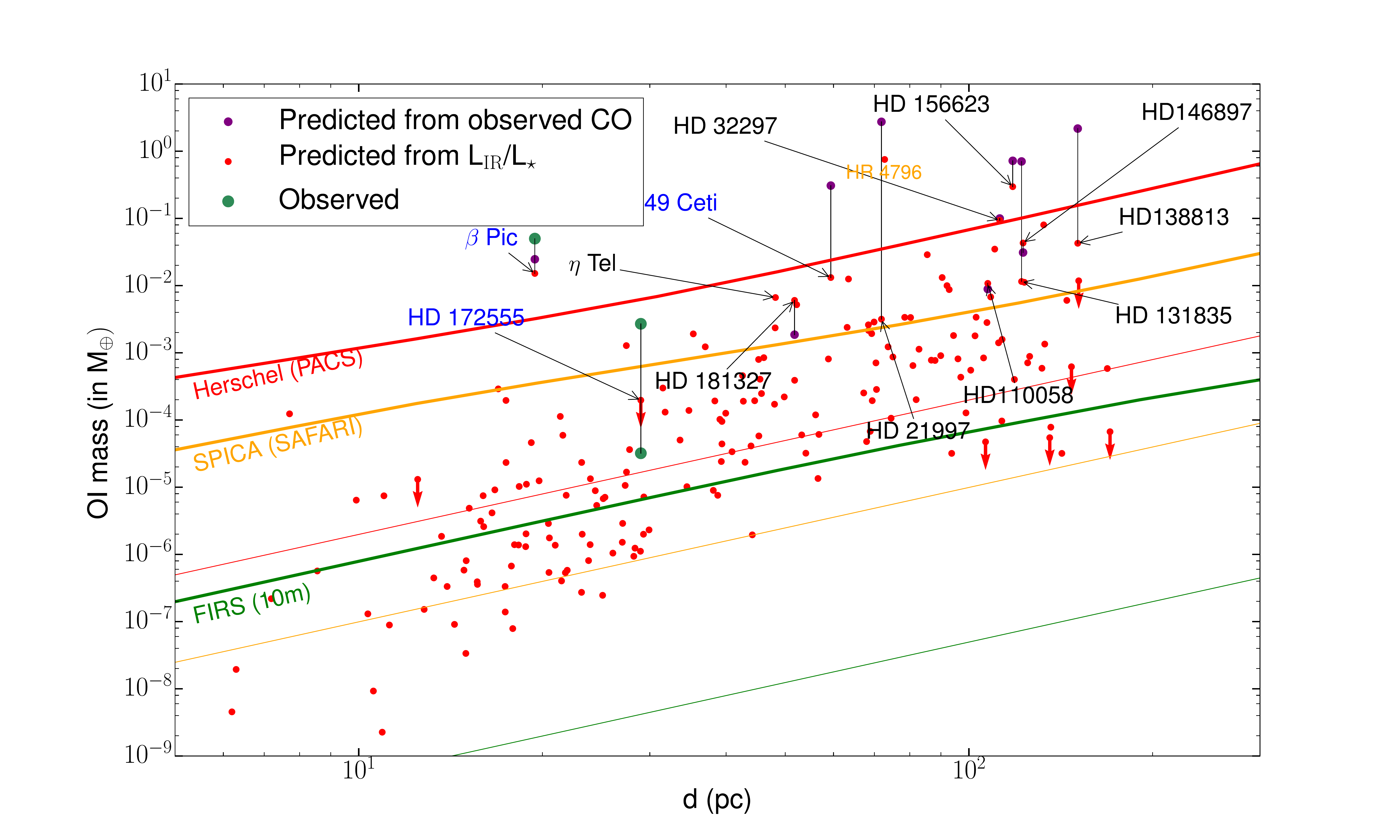}
   \caption{\label{figo1ap} OI mass (in M$_\oplus$) as a function of distance to Earth ($d$) when water + CO are released from solid bodies. Planetary systems with gas detections are labelled with their names. If OI is detected, the label is in blue (black otherwise). The OI mass for $\beta$ Pic, 
 derived from PACS observations \citep{2016MNRAS.461..845K,2016A&A...591A..27B} is shown as a green point as well as the observed mass for HD 172555 considering the flux observed with PACS \citep{2012A&A...546L...8R}. The second green point (at a lower mass) for HD 172555
considers that some extra hydrogen comes off the grains at the same time as CO is released (see section \ref{disc}).
The red points are predictions from our model. The red downward arrows show systems that are in the blue hatched area on Fig.~\ref{fig1}, which cannot keep CO trapped on solid bodies. The purple points show predictions from our model when the observed CO mass
is used rather than the CO mass predicted from $L_{\rm IR}/L_\star$. Detection limits at 5$\sigma$ in one hour are shown for Herschel/PACS (in red), SPICA/SAFARI (in orange) and FIRS (in green) for a 10m aperture. The thin lines are for LTE calculations and thick lines
for more realistic NLTE calculations (using the same assumptions as described in section \ref{detect}).}
\end{figure*}

We compute new predictions for OI assuming that the (CO+CO$_2$)/H$_2$O abundance ratio is $\sim$ 30\% \citep[i.e. an average Solar System composition, 2-60\%,][]{2011ARA&A..49..471M}. Therefore, the new oxygen mass released is $\sim$ 3 times higher than without extra water.
The new OI predictions are plotted in Fig.~\ref{figo1ap} and the most favourable targets are listed in Table \ref{tabo1b}. We see that the new prediction for $\beta$ Pic is closer to the observation. Also, SPICA would be able to detect OI in more systems due to both higher OI masses and higher excitation of the OI energy levels. Indeed, in this case,
we computed the NLTE lines taking into account the extra HI colliders that will further excite the OI energy levels. We assume that the ionization fraction is 0.1 so the corresponding $n_H/n_e \sim 60$ is input in the NLTE code. In this case the detection thresholds go
down and it becomes easier to detect OI as it is more excited. If the amount of water released is close to that assumed, SPICA could detect $\sim$ 30 systems with OI (see Table \ref{tabo1b} or Appendix \ref{tables} to get the full list) instead of 3 for the case without extra water.

When adding extra water in HD 172555, our flux prediction comes closer to the observed flux (see section \ref{ox} for the calculations without extra water). This can already be seen in Fig.~\ref{figo1ap}, where we 
computed the mass from the observed flux by PACS \citep{2012A&A...546L...8R} taking into account some extra hydrogen
in the same proportion as given in the previous paragraph. This new mass is much smaller than when there is no water ($3 \times 10^{-5}$ M$_\oplus$) but will produce the observed flux as it is more excited (by HI).
Our prediction for HD 172555 when extra water is released provides a total OI mass of 2.0$\times 10^{-4}$ M$_\oplus$ and a flux of 2.5$\times 10^{-17}$ W/m$^2$ which is close to the observed flux.
We find that assuming a standard CO/H$_2$O abundance ratio consistent with the composition of Solar System's comets is able to explain the observed flux and we then predict that the OI mass in HD 172555 is $\sim 10^{-4}$ M$_\oplus$. 
A better knowledge of the position of the gas in this system could help to know whether water can still remain on these warm grains. 

Interestingly, new missions such as SPICA could assess 
the amount of water released together with CO (as was done in KWC16 using the OI line). We notice that the more colliders, the more the NLTE lines get closer to the LTE regime, which is favourable for detections. However, in Fig.~\ref{figo1ap} the NLTE line is still 3 orders of magnitude above LTE even when 
assuming a Solar System comet composition to derive the amount of extra water released. Releasing water together with CO will not affect our predictions for CI, CII and CO masses but it may slightly change the NLTE lines that are plotted on the corresponding figures. It would make systems that are in NLTE
easier to detect. We concluded in KWC16 that the hydrogen was not playing any role in the thermal bugdet of the gas disc and we assume the same here, i.e. that the temperature will not vary when adding more water in the system. 


{\it Mass predictions:} The predictions presented in this paper are not guaranteed to exactly fit all new coming observations as there are some uncertain free parameters. However,
our default set of parameters still allows our model to work for a wide range of systems. For instance, we fixed the planetesimal eccentricity to be 0.05 but from one system to another, this could easily vary from 0.01 to 0.2. The masses predicted
vary as $e^{5/3}$ and could explain part of the differences with observations. For an individual system, one can refine the estimates for the free parameters. For instance, for a resolved belt, its width ${\rm d}r$ is known better than
within a factor 15 (that was assumed here in our error calculations). Therefore, our results should be used with caution (large error bars should be given) when dealing with the fiducial values rather than the specific parameters derived from a given system. However, all general results given here can be used as
a guide to further our understanding of gas and increase the amount of actual detections.  

{\it Flux predictions:} The flux predictions given in this paper should also be taken with caution. Our model computes a prediction for the mass that has to be converted to a flux. For CO, where the CMB dominates the excitation, the conversion is straighforward when the line is optically thin. However,
when lines become optically thick, the predicted flux will depend on optical thickness $\tau_\nu$. To compute $\tau_\nu$, we here assume that CO is in a ring centered at $R_0$ with a width ${\rm d}r/r=0.5$ and that the surface density is constant. This could be refined for some systems in the future when
ALMA provides resolved CO maps. Also, we assume an edge-on configuration when computing $\tau_\nu$, which is the most constraining case and so some of our flux predictions could be slightly underestimated.

The conversion from mass to fluxes for CI, CII and OI is more complicated and thus more uncertain. On top of the optical thickness assumptions cited above, the dust radiation field matters. The CMB at these shorter wavelengths is not dominant and it is mainly the dust radiation field
which drives the excitation of the lines. The dust radiation field that we know best at these wavelengths is for $\beta$ Pic. Thus, we assume that dust radiation field and scale it up or down for the different systems by comparing the flux found by fitting an SED to the $\beta$ Pic flux. 
We thus do not expect that our flux predictions match exactly future observations but they rather give an order of magnitude estimate to check which systems are most likely to be detectable. One should not rule out a target with a slightly lower predicted flux as this could go up if
we underestimated the dust radiation field impinging on the gas disc.

{\it Value of the viscosity parameter $\alpha$:} The atomic mass predicted will vary depending on $\alpha$. Throughout this paper, we assumed that $\alpha=0.5$ as predicted for the gas disc around $\beta$ Pic. $\alpha$ could be smaller around less ionized systems
and our mass predictions for CI or CII could go up by a factor $0.5/\alpha$. From our results, $\alpha$ is constrained to be relatively large for systems with high atomic masses as otherwise these could have been
detected with PACS (for CII) or APEX (for CI). However, for systems lying at the bottom of our plots, one cannot rule out that $\alpha$ is smaller. The mean free path of a gas atom goes up and one might reach another regime to transport angular momentum. Thus our atomic mass predictions
could go up and increase the numbers predicted for discoveries by ALMA, SPICA and FIRS.

If the sample with CO+CI gas detected grows significantly thanks to ALMA, and some free parameters can be refined (e.g. d$r$, ionization fraction, T$_{\rm gas}$, ...) with these same
observations, our model will lead to an estimate for $\alpha$. It could also show how $\alpha$ varies with ionization fraction. These new $\alpha$ estimates could then be compared to the most promising 
models at transporting angular momentum in discs, such as the MRI \citep[e.g.][]{2016MNRAS.461.1614K}.

{\it Detectability:} We emphasise that our predictions for detectability are to be taken as a guide rather than a fixed threshold. Indeed, the NLTE lines depend on the electron density, which vary with radial location. We picked $R_0=85$au in our study to plot the different lines but this should be
updated when targeting a specific system, or one should rather look at the flux predictions bearing in mind the caveats described above concerning these fluxes. Also, the optical thickness that is predicted depends on the system's geometry. Thus, these predictions should be made for each system individually when computing, for instance,
 the required total integration time to detect a specific transition of a specific species. Finally, the fluxes we provide in Appendix~\ref{tables} are also to be taken with caution. The outcome of our model are masses and we have to convert to a flux making some assumptions, as explained
in the flux prediction caveat paragraph above.

\subsection{Link to future observations}

The spatial distribution of the gas compared to the dust is very important to distinguish 
between different gas release mechanisms (since gas and dust will be colocated or at different 
positions for different scenarios), making ALMA the perfect tool to understand the origin 
of the gas.

One of the main outcomes of this study is that we find that ALMA could detect CI gas around at least 30 systems (and $\sim$ 15 with CO).
Using ALMA's high-resolution to resolve gas in inner regions of planetary systems could reveal some hidden components of planetary systems. It would enable us to probe the inner parts of planetary systems (for the brightest systems that can be spatially resolved) 
in a way that cannot be done using dust observations because the dust
is located farther from the central star. This could, for instance, enable to resolve structures in gas discs that are created by giant planets located in the inner regions.

We predict that OI observations with SPICA will give only a handful of detections if no water is released in the process of releasing CO, or $\sim$ 30 detections otherwise. However, having CO, CI or CII and OI detections can lead to predictions of the amount of water in exocomets 
and the amount of hydrogen in the gas phase as we have already shown for $\beta$ Pic in KWC16. Our prediction from KWC16 that hydrogen (together with carbon and oxygen) should be accreted on $\beta$ Pic has just been confirmed observationally by \citet{2016arXiv161200848W}.
Our model could be used to make predictions on the detectability of HI around other systems than $\beta$ Pic.
Probing the composition of exocomets and the amount of hydrogen observationally using the OI line provides motivation for a sensitive far-IR mission (such as FIRS). This would lead to the first extensive taxonomy of exocomets.

We find that only a small CO input rate ($>10^{-4}$ M$_\oplus$/Myrs) is sufficient to prevent CI from being pushed away by radiation pressure. However,
we note that the implications of this depend on the formation history of the gas in the system. Indeed, if there were no shielding from the very beginning, the CI or CII gas disc that would be building up over a viscous timescale would be blown out before reaching a sufficient amount of self-shielding. One can
imagine that these secondary discs are born at the end of the protoplanetary disc phase and that there were already some shielding from the star at that stage to prevent carbon from being blown out \citep{2015Ap&SS.357..103W}. It is not clear yet which is the right scenario and it might be that both scenarios can be found around
different systems and explain why some early type stars would have carbon observed and some others not. HR 4796 could be a case where a gas disc was never optically thick enough to prevent carbon gas from being blown out. However, the narrowness of the disc could be explained
if gas was present \citep{2001ApJ...557..990T}. New observations with APEX or ALMA of the CI line could distinguish between these two scenarios. 

We here suggest a new method to distinguish between a primordial versus secondary origin. Indeed, if our model cannot reproduce observations even assuming extreme values for our free parameters, we claim that the specific systems are likely to have a primordial origin. In this paper,
we show that this is the case for HD 21997, HD 131835 and HD 138813. It is complementary to other methods such as
observing optically thin line ratios to check whether there is enough molecular hydrogen around to be in LTE \citep{2016MNRAS}.

\section{Conclusion}\label{ccl}
We tested our new gas model developed for $\beta$ Pic in KWC16 on all systems with gas detected to check whether our model could explain all observations so far and then give predictions concerning future observations. 
The model assumes that CO gas observed around debris disc stars is secondary and is created from the solid volatile-rich bodies residing in the parent belt of the discs. Once CO gas is created, it photodissociates into carbon and oxygen atoms, which viscously spread to form
an atomic accretion disc inside the parent belt and a decretion disc outside. The model calculates the ionization fraction of carbon, the gas temperature and population levels at different radial locations in the disc. It also takes into account CO self-shielding against photodissociation and
CI self-shielding against photoionization. When computing the detectability, we take into account NLTE effects and optical thickness of lines.

We find that our model is able to explain most current observations. Systems that we predict to be detectable are indeed detected and systems that lie under the detection threshold are not. Only for 3 systems our model cannot reproduce observations. 
We suggest that these 3 systems, HD 21997, HD 131835 and HD 138813 are not made only of secondary gas but still possess some primordial gas. In this sense, our model can rule out a secondary origin for some discs, suggesting rather a primordial origin. 

We provide an analytical formulation of our model through a set of equations in this paper. We clearly identify the most important parameters that lead to enhance CO, CI, CII or OI abundances. We define some regions of the parameter space in terms of star's luminosity versus planetesimal belt location
where CO is not expected to be observed. For instance, if grains are too warm ($\gtrsim$ 140K), we do not expect CO to be retained on grains (e.g. HD 172555) or if the star is too luminous, the CO
photodissociation timescale gets too short to detect CO (e.g. $\eta$ Tel). We find
that we should observe systems with the highest fractional luminosities, closest to Earth and having L$_\star$ small enough and the parent belt radial location far enough not to lie in the hatched exclusion areas defined in Fig~\ref{fig1}. 
Also, we study the effect of a change in the assumed fiducial parameter values and give the maximum variations expected for each parameter.   

Someone wanting to make predictions for the amount of CO in a specific system can use Eq.~\ref{mco} (that uses Eqs~\ref{massloss} and \ref{COloss}). The CO photodissociation timescale can be assumed to be 120 years for systems with debris far from their host star, with the latter
not being too luminous (see Fig.~\ref{fig1}). The self-shielding factor $\epsilon_{\rm CO}$ can be computed using Fig.~\ref{figmcomdot}. To derive the total carbon mass or OI mass (assuming no extra water), one should use Eq.~\ref{carbmass}. If extra water is assumed to be released together
with CO, the OI mass will increase and depends on the assumed H/C ratio. Eqs.~\ref{mc2} and \ref{mc1} can be used to predict the CII and CI masses in a specific system, where the carbon ionization fraction $f$ can be computed using Eq.~\ref{ionfrac} (or assumed to have a typical value of 0.1
for a first guess).

Based on these results, we can use our model to make predictions for systems that have no gas detected so far. To this aim, we have taken a sample of 189 debris disc stars and ran our model for each of them. We make predictions for new detections with ALMA, and with potential future
missions such as SPICA and a far-IR 10m telescope (e.g. FIRS). We predict that ALMA could detect at least 15 systems with CO and 30 with CI (with less than an hour of integration time for each target). CI seems the most promising avenue for the near future and one could use
ALMA's high resolution to probe the inner regions of planetary systems through CI that extends all the way to the star (accretion disc), which may indirectly reveal some hidden planets.
SPICA will enable us to detect at least 25 new systems with CII and $\sim$ 30
with OI (depending on the amount of water released together with CO). To detect OI around a fair number of stars, a new far-IR 10m telescope (such as FIRS) is needed. We give a list of the systems that are most likely to be detected with ALMA in CO and CI in Tables \ref{tab2} and \ref{tab3} and
with SPICA in CII and OI in Tables \ref{tabc2} and \ref{tabo1b}.

We also recomputed the OI mass in HD 172555 with our NLTE model from the observed flux with PACS and find $3 \times 10^{-5}$ M$_\oplus$, which can be explained with our second generation gas model.

We find that CO, CI, CII and OI gas should be modelled in non-LTE for almost all systems, and for the most gas-rich debris discs, CO, CI and OI lines will be optically thick.

In this paper, we also study the effect of radiation pressure on carbon and oxygen. Around luminous stars, CI is expected to be blown out. We find that a small CO input rate ($\sim$ 1000 times smaller than in $\beta$ Pic) is enough to create a shielding from the star that significantly reduces radiation pressure
and allows for CI to stay bound. Therefore, our model explains self-consistently for the first time why carbon was detected around early type stars.

\section*{Acknowledgments}
We thank the referee for his/her detailed review.
QK, LM and MCW acknowledge support from the European Union through ERC grant number 279973. GMK is supported by the Royal Society as a Royal Society University Research Fellow. 
{\it Herschel} is an ESA space observatory with science instruments provided by European-led Principal Investigator consortia and with important participation from NASA.

\appendix

\section{Calculations of the Temperature in our model}\label{tempcalc}

The main heating mechanism in our model is assumed to be photoionization of carbon (see KWC16). Each ionized neutral will give away an electron that will contribute to heating the gas. The amount of energy given to the gas is the difference between the initial photon energy and the ionization potential needed to get ionized.
Therefore, the total rate of energy created per unit volume due to photoionization is

\begin{equation}
\Gamma_{\rm ion}=n_{\rm C_{I}} \int_{\nu_{\rm IP}}^\infty \frac{4\pi J_\nu}{h \nu} h(\nu -\nu_{\rm IP}) \sigma_{\rm ion}(\nu) \, {\rm d}\nu,
\end{equation}

\noindent where $\nu_{\rm IP}=c/\lambda_{\rm IP}$ is the minimum frequency to get ionized. $\lambda_{\rm IP}=1100$ \AA, which corresponds to an ionization potential of 11.26eV. For carbon, the ionization cross section does not vary with wavelength and is equal to 
$\sigma_{\rm ion}(\nu)=1.6\times 10^{-17}$ cm$^2$ between 912 (Lyman break) and 1100 \AA\,\citep{2008CP....343..292V}.

The main coolant in our model is assumed to be the CII fine structure line at 157.7$\mu$m (see KWC16). Cooling by the OI fine structure line at 63.2$\mu$m can also be important. However, when carbon is present, and for the range of temperatures expected in gas discs, the CII line is always the dominant coolant \citep{2010ApJ...720..923Z}.
Statistical equilibrium requires that the upward transitions (collisional excitations) balance the downwards ones (collisional de-excitations and spontaneous radiative decays). For a two-level atom model that represents well the CII line, 
it gives \citep[following][]{2010ApJ...720..923Z}

\begin{equation}
\label{eqeq}
n_e n_1 q_{1,2}=n_e n_2 q_{2,1}+n_2 A_{2,1},
\end{equation}

\noindent where $n_e$, $n_1$ and $n_2$ are the number densities of electrons and atoms in the lower and upper levels, respectively. $q_{1,2}$ and $q_{2,1}$ are the collisional excitation and de-excitation rates and $A_{2,1}$ is the Einstein coefficient. The gas cooling rate is then defined as \citep{2010ApJ...720..923Z}

\begin{equation}
\Lambda_{1,2}=(n_2 q_{2,1} - n_1 q_{1,2}) \, n_e h \nu_{1,2},
\end{equation}

\noindent which can be expressed in terms of CII number density $n_{\rm CII}=n_1+n_2$ using Eq.~\ref{eqeq}

\begin{multline}
 \Lambda_{1,2} = \xi \left( 1 - \frac{A_{2,1}+n_e q_{2,1}}{A_{2,1}+n_e (q_{2,1}+q_{1,2})} \right. \\
\left. \times (1 + \frac{g_2}{g_1} e^{-h \nu_{1,2}/k_B T_{\rm gas}}) \right),
\label{lambda12}
\end{multline}

\noindent where $g_1$ and $g_2$ are the statistical weight of the lower and upper level, $\xi= q_{2,1} n_e n_{\rm CII} h \nu_{1,2}$, and $q_{1,2}=q_{2,1} e^{-h \nu_{1,2}/k_B T_{\rm gas}} $.


We can then solve for $T_{\rm gas}$ equating $\Gamma_{\rm ion}$ to $\Lambda_{1,2}$ to get the temperature as a function of the radial position in the gas disc. We applied this method and compare with the numerically calculated temperature profile (using CLOUDY) for $\beta$ Pictoris.

\begin{figure}
   \centering
   \includegraphics[width=8.5cm]{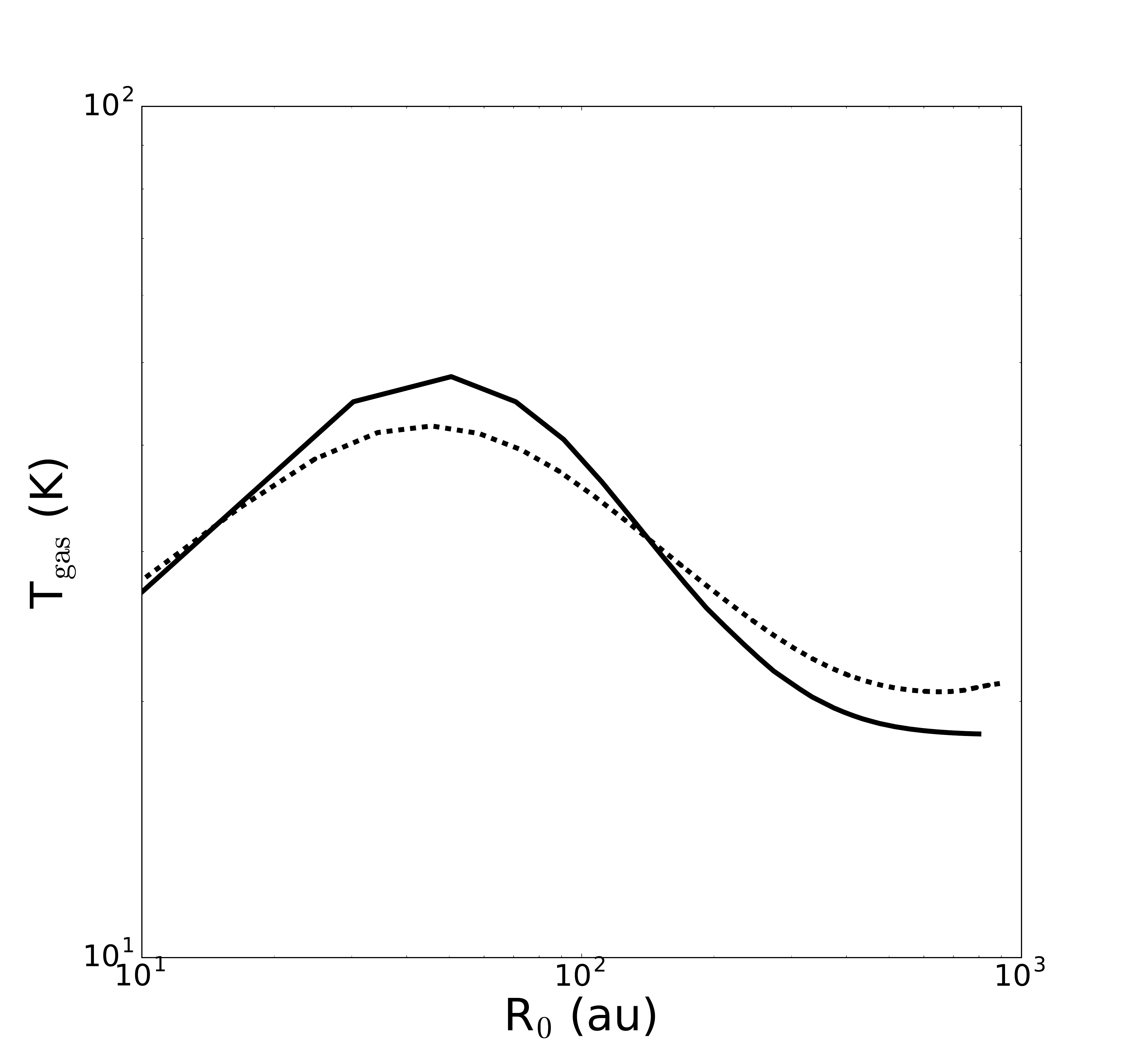}
   \caption{\label{figappen} Gas temperature radial profile predicted with our analytical model (solid line) and from the numerical model CLOUDY (dashed line).}
\end{figure}

In KWC16, the best-fit model was for an interstellar radiation field that was $\sim$60 times the standard value. To have a meaningful comparison, we do not use the best-fit model but rather one of the other models with a standard IRF presented in KWC16. 
In Fig.~\ref{figappen}, we plot the gas temperature for this case coming from a CLOUDY numerical simulation (dashed line) and compare it to the gas profile we obtain from our analytical model (solid line). We find that our analytical model well reproduces
the overall shape of the gas profile.

\section{NLTE calculations}\label{NLTE}

Deriving a total gas mass from an observed integrated line flux for a particular transition of a given species requires knowledge of the fractional population of the upper energy level of the transition. In other words, we need to know the fraction of mass populating the upper level in question relative
 to the total mass of the species.
Non-local thermodynamic equilibrium treatment of the excitation of gas species is crucial for calculating these fractional level populations in low density astrophysical environments, where the density of collisional partners is likely to drop below the critical density necessary for the LTE
 approximation to be valid. Here, based on the formalism developed in \citet{2015MNRAS.447.3936M} for the CO molecule, we extend our excitation code to solve the full NLTE statistical equilibrium for atomic species OI, CI and CII. Given the local radiation field $J_{\nu_{ul}}$ at the frequency of each transition
 between any two upper ($u$) and lower ($l$) levels, the density of main collisional partners $n_{\rm coll}$ and the kinetic temperature of the gas $T_{\rm kin}$, the code solves the statistical equilibrium and outputs the fractional population $x_i$ of all energy levels $i$ of the species
 considered. We direct the reader to Sect. 2 in \citet{2015MNRAS.447.3936M} for a more extensive description of the method and the theory behind it.

\begin{figure*}
   \centering
   \includegraphics[width=16.5cm]{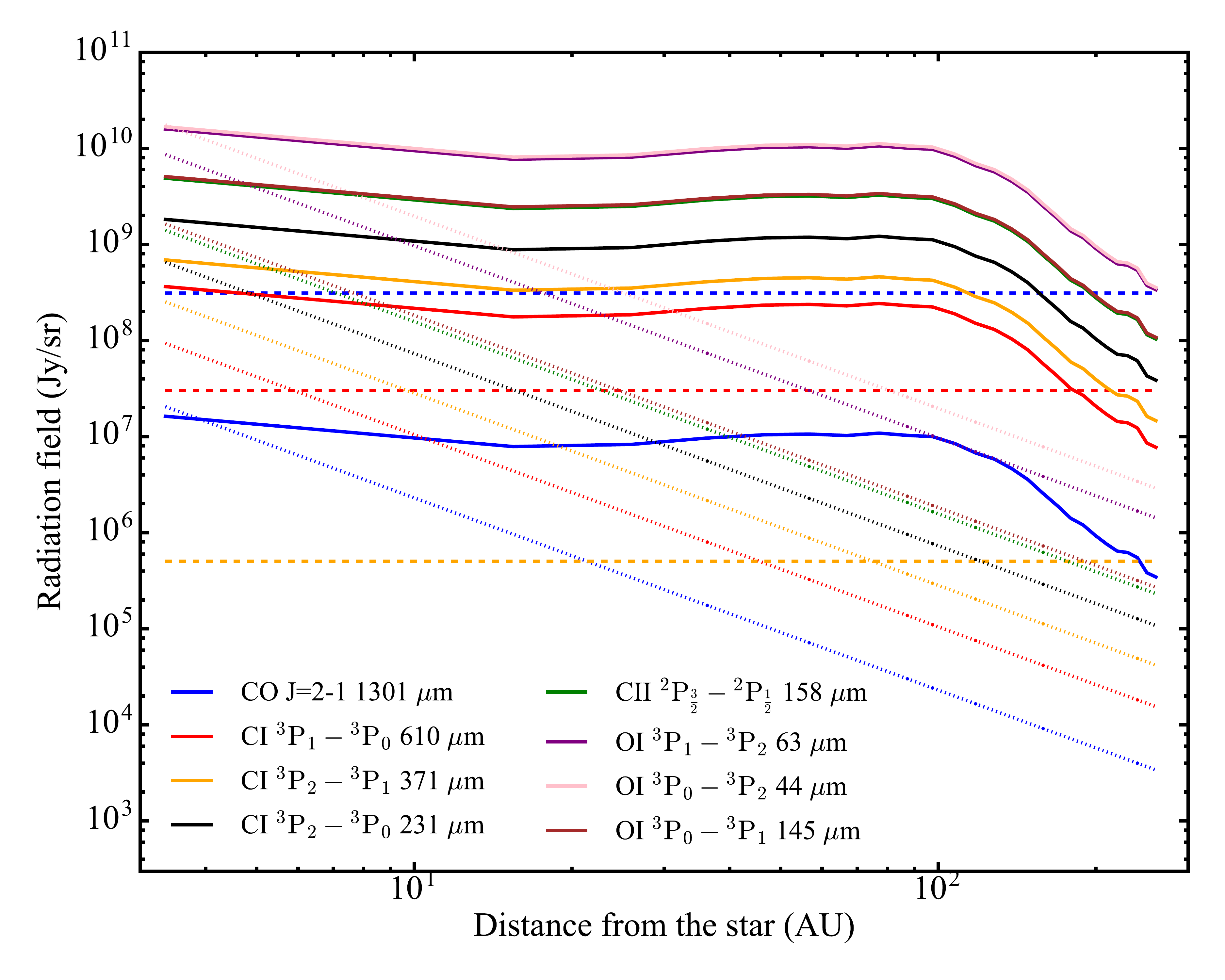}
   \caption{\label{figrad} Radiation field impinging for different species assuming a $\beta$ Pic like dust and star emission. Solid lines are for dust radiation, dashed for the CMB and dotted for star's radiation at the transition wavelength.}
\end{figure*}

For each species, we obtain energy levels, transition frequencies and Einstein coefficients from the Leiden Atomic and Molecular DAtabase \citep[LAMDA,][]{2005A&A...432..369S}. In addition, we assume electrons released from carbon photoionization to be the dominant collisional partner (see Discussion), 
and obtain collisional rate coefficients from the same database. 
The radiation field $J_{\nu_{ul}}$ at the wavelengths of each transition between any two levels is made up of 3 contributions; the cosmic microwave background (CMB), the dust emission, and the stellar emission. The stellar emission is calculated using stellar models matching 
the spectral type of the star as described in Sect. 3.2, whereas the radiation field due to dust emission is calculated using the RADMC-3D code\footnote{\url{http://www.ita.uni-heidelberg.de/ dullemond/software/radmc-3d}} for the best-fit model to high-resolution 1.3 millimetre
 observations of the $\beta$ Pictoris disc (Matr\`a et al. in prep.). We then assume that the spatial distribution of dust emission is independent of wavelength (i.e. there is no spatial segregation of grains), and calculate the radiation field at other wavelengths by simply scaling
 it using fluxes from the SED \citep{2014MNRAS.444.3164K}. For other stars, we scale this radiation field up or down to match the total intensity predicted from the SED of the star at the given wavelength. Fig.~\ref{figrad} shows an example comparison of the stellar, CMB and dust 
contributions for the $\beta$ Pictoris disc as a function of distance from the central star, for transitions of the species relevant to this work. As expected, we see that the CMB dominates the radiation field at long mm wavelengths, with the dust contribution from the $\beta$ Pic
 belt becoming more important already at 610 $\mu$m and shorter wavelengths. The stellar contribution, on the other hand, only becomes comparable to that of the dust within a few AU from the star.

Given this total radiation field for all transitions, the assumed electron density and temperature as discussed in Sect. 3.5.1 and 3.5.3, we solve the full NLTE statistical equilibrium for OI, CI and CII to derive the fractional level population $x_i$ for all levels $i$ of each species. 
For a given observed transition, the upper level fractional population can then be used through Eq.~\ref{lineflux} to derive a gas mass from an observed flux, and vice versa.





\section{Tables describing our sample of 189 stars and giving our predictions for each star}\label{tables}

\xentrystretch{-0.1}

\clearpage

\afterpage{
\onecolumn
Description of the 189 stars used in this study. {\it Column 1:} Star's name. {\it Column 2:} Distance to Earth (in pc). {\it Column 3:} Star's temperature (in K). {\it Column 4:} Star's Luminosity (in L$_\odot$). {\it Column 5:} Dust fractional luminosity. {\it Column 6:} Location of the belt (au). {\it Column 7,8,9:} Flux at 60, 160, 610 microns (in W/m$^2$).\\
\tablefirsthead{\hline Name & $d$  & $T_{\rm eff}$ &  $L_\star$ & $L_{\rm IR}/L_\star$  &  $R_0$& $F_{\rm 60}$  &  $F_{\rm 160}$  & $F_{\rm 610}$ \\
 & (pc)  & (K) & ($L_\odot$) &  & (au) &  (Jy) & (Jy) & (Jy) \\ \hline}
\tablehead{\hline
\multicolumn{9}{|l|}{\small\sl continued from previous page}\\ \hline Name & $d$  & $T_{\rm eff}$ &  $L_\star$ & $L_{\rm IR}/L_\star$  &  $R_0$& $F_{\rm 60}$  &  $F_{\rm 160}$  & $F_{\rm 610}$ \\
 & (pc)  & (K) & ($L_\odot$) &  & (au) &  (Jy) & (Jy) & (Jy) \\ \hline} 
\begin{xtabular*}{18.5cm}{@{\extracolsep{\fill}}|*{9}{p{2.cm}|p{1.2cm}|p{1.2cm}|p{1.2cm}|p{1.9cm}|p{1.2cm}|p{2.cm}|p{2.cm}|p{2.cm}|}}

$\beta$ Pic & 19.4 & 7405 & 8.7 & 1.7$\times 10^{-03}$ & 85 & 1.3$\times 10^{+01}$ & 4.1$\times 10^{+00}$ & 1.7$\times 10^{-01}$ \\
$\eta$ Tel & 48.2 & 9379 & 22 & 7.6$\times 10^{-04}$ & 24 & 4.7$\times 10^{-01}$ & 7.6$\times 10^{-02}$ & 2.0$\times 10^{-03}$ \\
49 Ceti & 59.4 & 8922 & 15.5 & 1.1$\times 10^{-03}$ & 100 & 1.8$\times 10^{+00}$ & 1.1$\times 10^{+00}$ & 3.7$\times 10^{-02}$ \\
HD 110058 & 107 & 8006 & 5.9 & 1.9$\times 10^{-03}$ & 50 & 3.7$\times 10^{-01}$ & 4.8$\times 10^{-02}$ & 1.2$\times 10^{-03}$ \\
HD 131835 & 122 & 7984 & 9.2 & 1.5$\times 10^{-03}$ & 50 & 6.3$\times 10^{-01}$ & 3.7$\times 10^{-01}$ & 2.3$\times 10^{-02}$ \\
HD 138813 & 150.8 & 8941 & 24.5 & 1.5$\times 10^{-04}$ & 100 & 4.1$\times 10^{-01}$ & 6.7$\times 10^{-02}$ & 1.7$\times 10^{-03}$ \\
HD 146897 & 122.7 & 6256 & 3.1 & 5.4$\times 10^{-03}$ & 100 & 6.6$\times 10^{-01}$ & 1.0$\times 10^{-01}$ & 2.8$\times 10^{-03}$ \\
HD 156623 & 118 & 8576 & 14.8 & 5.5$\times 10^{-03}$ & 75 & 9.8$\times 10^{-01}$ & 1.1$\times 10^{-01}$ & 2.7$\times 10^{-03}$ \\
\hline
\end{xtabular*}

\clearpage
\vspace{2cm}

Model predictions for the 189 stars used in this study (masses are in M$_\oplus$ and fluxes in W/m$^2$). {\it Column 1:} Star's name. {\it Column 2:} CO mass. {\it Column 3,4:} CO flux at 1.3mm, 870microns. {\it Column 5:} CI mass. {\it Column 6:} CI flux at 610microns. {\it Column 7:} CII mass. {\it Column 8:} CII flux at 158microns. {\it Column 9:} OI mass (with extra water, see Sect.~\ref{cavea}). {\it Column 10:} OI flux at 63microns (with extra water).\\
\tablefirsthead{\hline Name & $M_{\rm CO}$  & $F_{\rm CO}$ \tiny{1300} &  $F_{\rm CO}$ \tiny{870} & $M_{\rm CI}$  &  $F_{\rm CI}$ \tiny{610}& $M_{\rm CII}$  &  $F_{\rm CII}$ \tiny{158} & $M_{\rm OI}$  & $F_{\rm OI}$ \tiny{63}\\
 & ($M_\oplus$)  & (W/m$^2$) & (W/m$^2$) & ($M_\oplus$)  & (W/m$^2$) &  ($M_\oplus$) & (W/m$^2$) & ($M_\oplus$) & (W/m$^2$) \\ \hline}
\tablehead{\hline
\multicolumn{10}{|l|}{\small\sl continued from previous page}\\ \hline Name & $M_{\rm CO}$  & $F_{\rm CO}$ \tiny{1300} &  $F_{\rm CO}$ \tiny{870} & $M_{\rm CI}$  &  $F_{\rm CI}$ \tiny{610}& $M_{\rm CII}$  &  $F_{\rm CII}$ \tiny{158} & $M_{\rm OI}$  & $F_{\rm OI}$ \tiny{63}\\
 & ($M_\oplus$)  & (W/m$^2$) & (W/m$^2$) & ($M_\oplus$)  & (W/m$^2$) &  ($M_\oplus$) & (W/m$^2$) & ($M_\oplus$) & (W/m$^2$) \\ \hline} 

\begin{xtabular*}{18.5cm}{@{\extracolsep{\fill}}|*{10}{p{2.cm}|p{1.35cm}|p{1.35cm}|p{1.35cm}|p{1.35cm}|p{1.35cm}|p{1.35cm}|p{1.35cm}|p{1.35cm}|p{1.35cm}|}}

$\beta$ Pic & 4.6$\times 10^{-05}$ & 2.4$\times 10^{-20}$ & 1.9$\times 10^{-20}$ & 3.4$\times 10^{-03}$ & 2.3$\times 10^{-19}$ & 3.8$\times 10^{-04}$ & 6.9$\times 10^{-18}$ & 1.5$\times 10^{-02}$ & 1.4$\times 10^{-17}$ \\
$\eta$ Tel & 8.5$\times 10^{-08}$ & 1.1$\times 10^{-23}$ & 4.9$\times 10^{-23}$ & 1.1$\times 10^{-03}$ & 2.1$\times 10^{-20}$ & 5.6$\times 10^{-04}$ & 4.0$\times 10^{-19}$ & 6.6$\times 10^{-03}$ & 1.6$\times 10^{-18}$ \\
49 Ceti & 4.0$\times 10^{-06}$ & 3.6$\times 10^{-22}$ & 3.7$\times 10^{-22}$ & 2.2$\times 10^{-03}$ & 4.9$\times 10^{-20}$ & 1.0$\times 10^{-03}$ & 1.8$\times 10^{-18}$ & 1.3$\times 10^{-02}$ & 3.2$\times 10^{-18}$ \\
HD 110058 & 1.9$\times 10^{-04}$ & 9.5$\times 10^{-22}$ & 1.3$\times 10^{-21}$ & 2.6$\times 10^{-03}$ & 5.4$\times 10^{-21}$ & 1.3$\times 10^{-04}$ & 8.1$\times 10^{-20}$ & 1.1$\times 10^{-02}$ & 3.0$\times 10^{-19}$ \\
HD 131835 & 2.6$\times 10^{-05}$ & 5.6$\times 10^{-22}$ & 8.5$\times 10^{-22}$ & 2.7$\times 10^{-03}$ & 1.4$\times 10^{-20}$ & 1.7$\times 10^{-04}$ & 8.8$\times 10^{-20}$ & 1.2$\times 10^{-02}$ & 8.5$\times 10^{-19}$ \\
HD 138813 & 2.5$\times 10^{-06}$ & 4.2$\times 10^{-23}$ & 1.0$\times 10^{-22}$ & 6.7$\times 10^{-03}$ & 2.0$\times 10^{-20}$ & 4.0$\times 10^{-03}$ & 5.7$\times 10^{-19}$ & 4.3$\times 10^{-02}$ & 1.2$\times 10^{-18}$ \\
HD 146897 & 2.9$\times 10^{-03}$ & 1.1$\times 10^{-21}$ & 1.4$\times 10^{-21}$ & 1.0$\times 10^{-02}$ & 5.7$\times 10^{-21}$ & 4.6$\times 10^{-04}$ & 1.9$\times 10^{-19}$ & 4.3$\times 10^{-02}$ & 3.3$\times 10^{-19}$ \\
\end{xtabular*}
\twocolumn
}

\label{lastpage}

\end{document}